\documentclass[11pt]{paper}

\usepackage{apacite}
\usepackage[authoryear,round]{natbib}
\usepackage{url,hyperref}
\usepackage{nicefrac}
\usepackage{booktabs}
\usepackage{graphicx}
\usepackage{amsmath}

\usepackage{a4wide}
\usepackage{graphicx}
\usepackage{booktabs,array,multirow,xcolor,colortbl}

\begin{document}
\onecolumn
\graphicspath{{../img/}}
\title{Phylogenetic typology}

\author{Gerhard J\"ager \and Johannes Wahle\\
  University of T\"ubingen\
  Institute of Linguistics\\
  Wilhelmstr.\ 19, 72074 T\"ubingen, Germany\\
  Email: gerhard.jaeger@uni-tuebingen.de, johannes.wahle@uni-tuebingen.de}



\maketitle

\begin{abstract}
 In this article we propose a novel method to estimate the frequency distribution of linguistic variables while controlling for statistical non-independence due to shared ancestry. Unlike previous approaches, our technique uses all available data, from language families large and small as well as from isolates, while controlling for different degrees of relatedness on a continuous scale estimated from the data. Our approach involves three steps: First, distributions of phylogenies are inferred from lexical data. Second, these phylogenies are used as part of a statistical model to statistically estimate transition rates between parameter states. Finally, the long-term equilibrium of the resulting Markov process is computed.

 As a case study, we investigate a series of potential word-order correlations across the languages of the world.
\end{abstract}

\section{Introduction}

One of the central research topics of linguistic typology concerns the distribution of
structural properties across the languages of the world. Typologists are concerned with
describing these distributions, understanding their determinants and identifying possible
distributional dependencies between different linguistic features. Greenbergian language
universals \citep{greenberg63} provide prototypical examples of typological
generalizations. Absolute universals\footnote{E.g., Greenberg's Universal 1 ``In
 declarative sentences with nominal subject and object, the dominant order is almost
 always one in which the subject precedes the object.''} describe the distribution of a
single feature, while implicative universals\footnote{For instance Universal 3 ``Languages
 with dominant VSO order are always prepositional.''} state a dependency between
different features. In subsequent work (such as \citealt{dryer92}), the quest for
implicative universals was generalized to the study of \emph{correlations} between
features.

Validating such kind of findings requires statistical techniques, and the quest for
suitable methods has been a research topic since thirty years. A major obstacle is the
fact that languages are not independent samples --- pairwise similarities may be the
result of common descent or language contact. As the common statistical tests presuppose
independence of samples, they are not readily applicable to cross-linguistic data.

One way to mitigate this effect --- pioneered by \cite{bell78}, \cite{dryer89}, and
\cite{perkins89} --- is to control for genealogy and areal effects when sampling. In the
simplest case, only one language is sampled per genealogical unit, and statistical effects
are applied to different macro-areas independently. More recent work often uses more
sophisticated techniques such as repeated stratified random sampling (e.g.,
\citealt{blasietal16}). Another approach currently gaining traction is the usage of
(generalized) mixed-effects models \citep{breslowClayton1993}, where genealogical units
such as families or genera, as well as linguistic areas, are random effects (see, e.g.,
\citealt{atkinson2011,bentzWinter13,jaegeretal2011} for applications to typology).

In a seminal paper, \cite{maslova2000} proposes an entirely different conceptual take on
the problems of typological generalizations and typological sampling. Briefly put, if
languages of type A (e.g., nominative- accusative marking) are more frequent than
languages of type B (e.g., ergative-absolutive marking), this may be due to three
different reasons: (1) diachronic shifts B$\rightarrow$A are more likely than shifts
A$\rightarrow$B; (2) proto-languages of type A diversified stronger than those of type B,
and the daughter languages mostly preserve their ancestor's type, and (3) proto-world, or
the proto-languages at relevant prehistoric population bottlenecks, happened to be of type
A, and this asymmetry is maintained due to diachronic inertia. Only the first type of
reason is potentially linguistically interesting and amenable to a cognitive or functional
explanation. Reasons of category (2) or (3) reflect contingent accidents. Stratified
sampling controls for biases due to (2), but it is hard to factor (1) from (3) on the
basis of synchronic data. Maslova suggests that the theory of Markov processes offers a
way out. If it is possible to estimate the diachronic transition probabilities, and if one
assumes that language change has the Markov property (i.e., is memoryless), one can
compute the long-term equilibrium probability of this Markov process. This equilibrium
distribution should be used as the basis to identify linguistically meaningful
distributional universals.

\cite{maslovaNikitina2007} make proposals how to implement this research program
iinvolving the systematic gauging of the distribution of the features in question within
language families.

\cite{bickel2011,bickel2013} introduces the \emph{Family Bias Theory} as a statistical
technique to detect biased distributions of feature values across languages of different
lineages while controlling for statistical non-independence. Briefly put, the method
assesses the tendency for biased distributions within families on the basis of large
families, and extrapolates the results to small families and isolates.

In this article we will propose an implementation of Maslova's program that makes use of
algorithmic techniques from computational biology, especially the \emph{phylogenetic
 comparative method}. A technically similar approach has been pursued by
\cite{dunnetal11}, where it was confined to individual language families. In the present
paper we will propose an extension of their method that uses data from several language
families and isolates simultaneously. Unlike the above-mentioned approaches, our method
makes use of the entire phylogenetic structures of language families including branch
lengths --- to be estimated from lexical data ---, and it treats large families, small
families and isolates completely alike. Also, our method is formulated as a generative
model in the statistical sense. This affords the usage of standard techniques from
Bayesian statistics such as inferring posterior uncertainty of latent parameter values,
predictive checks via simulations, and statistical model comparison.

The model will be exemplified with a study of the potential correlations between eight
word-order features from the \emph{World Atlas of Language Structure} \citep{wals13} that
were also used by \cite{dunnetal11}.

\section{Statistical analysis}

\subsection{Continuous time Markov processes}

Following \cite{maslova2000}, we assume that the diachronic change of typological features
follows a continuous time Markov process (abbreviated as CTMC, for continuous time Markov
chain). Briefly but, this means that a language is always in one of finitely many
different states. Spontaneous mutations can occur at point in time. If a mutation occurs,
the language switches to some other state according to some probability distribution. This
process has the Markov property, i.e., the mutation probabilities --- both the chance of a
mutation occurring, and the probabilities of the mutation targets --- only depend on the
state the language is in, not on its history.

Mathematically, the properties of such a process can be succinctly expressed in a single
$n\times n$-matrix $Q$, where $n$ is the number of states. The entries along the diagonal
are non-positive and all other entries non-negative. Each row sums up to 0. The waiting
time until the next mutation, when being in state $i$, is an exponentially distributed
random variable with rate parameter $-q_{ii}$. If a mutation occurs while being in state
$i$, the probability of a mutation $i\rightarrow j$ is proportional to $q_{ij}$.

The probability of a language ending up in state $j$ after a time interval $t$ when being
in state $i$ at the beginning of the interval (regardless of the number and type of
mutations happening during the interval) is $p_{ij}$, where
$$
 \begin{aligned}
  P & = e^{tQ}.
 \end{aligned}
$$

According to the theory of Markov processes\footnote{See, e.g.,
 \citep{grimmettStirzaker}.}, if each state can be reached from each other state in a
finite amount of time, there is a unique equilibrium distribution $\pi$. Regardless of the
initial distribution, the probability of a language being in state $i$ after time $t$
converges to $\pi_i$ when $t$ grows to infinity. Also, the proportion of time a
language spent in state $i$ during a time interval $t$ converges to $\pi_i$ when $t$ grows
to infinity. According to Maslova, it is this equilibrium distribution $\pi$ that affords
linguistic insights and that therefore should be identified in distributional typology.\\

\subsection{Phylogenetic Markov chains}

\begin{figure}
 \begin{center}
  \includegraphics[width=7cm]{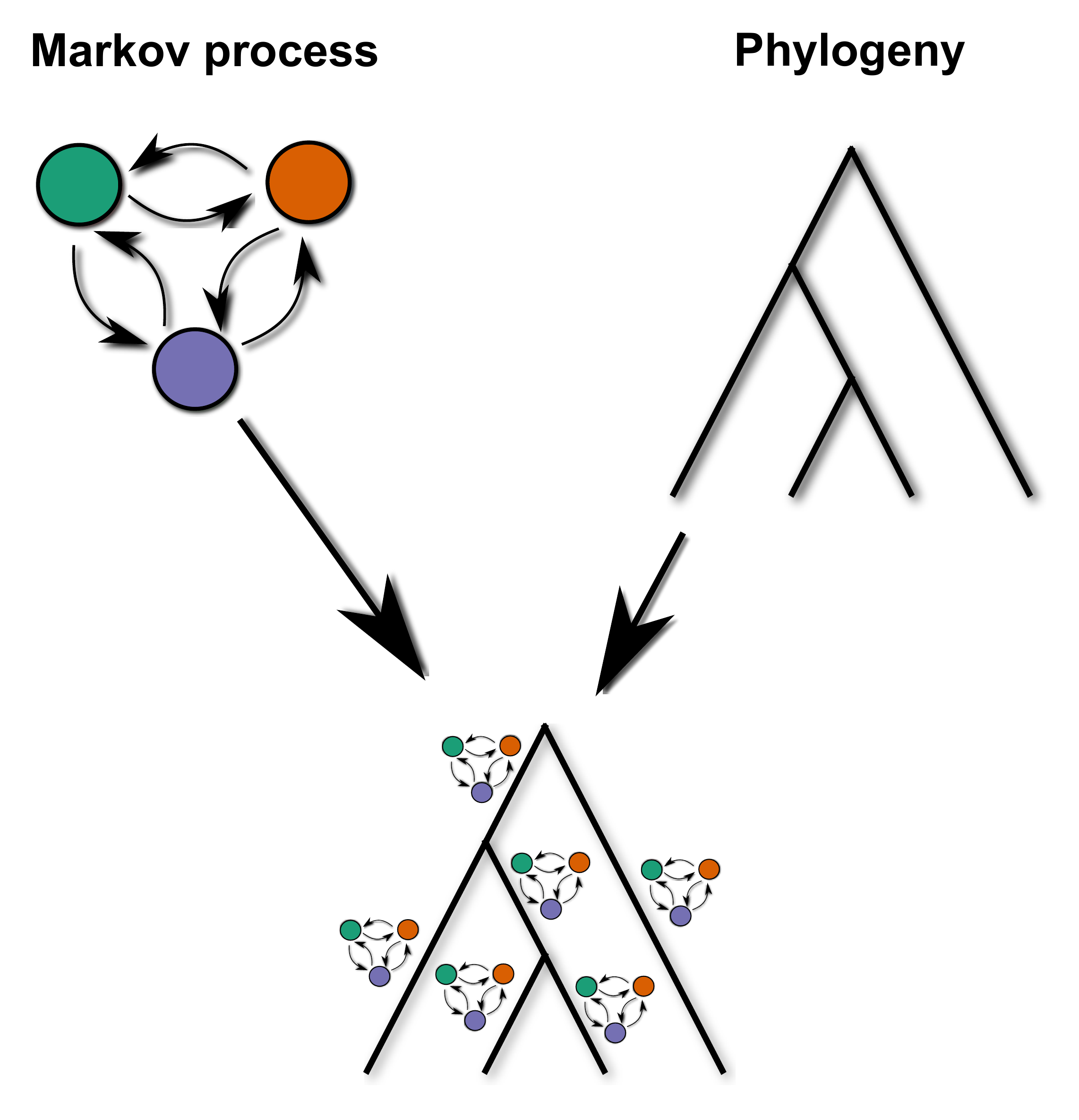}
 \end{center}
 \caption{ Schematic structure of the phylogenetic CTMC model. Independent but identical
  instances of a CTMC run on the branches of a phylogeny}\label{fig:1}
\end{figure}

Different languages are not samples from independent runs of such a CTMC. Rather, their
properties are correlated due to common descent, which can be represented by a
\emph{language phylogeny}. A phylogeny is a family tree of related languages with the
common ancestor at the root and the extant (or documented) languages at the leaves. Unlike
the family trees used in classical historical linguistics, branches of a phylogeny have a
length, i.e., a positive real number that is proportional to the time interval the branch
represents. According to the model used here, when a language splits in two daughter
languages, those initially have the same state but then evolve independently according to
the same CTMC. This is schematically illustrated in Figure \ref{fig:1}.

Let us illustrate this with an example. Suppose the feature in question has three possible
values, $a$, $b$ and $c$. The $Q$-matrix characterizing the CTMC is given in (\ref{eq:1}).
\begin{equation}
 \label{eq:1}
 \begin{aligned}
  Q & =
  \left(
  \begin{array}{rrr}
   -3 & 2  & 1  \\
   5  & -6 & 1  \\
   2  & 3  & -5
  \end{array}
  \right)
 \end{aligned}
\end{equation}

The equilibrium distribution\footnote{This is the left null vector of $Q$,
 normalized such that it sums to 1.} $\pi$ for this CTMC is
\begin{equation}
 \label{eq:2}
 \begin{aligned}
  \pi & = (\nicefrac{9}{16}, \nicefrac{13}{48}, \nicefrac{1}{6}) \\
      & \approx (0.56, 0.27, 0.17)
 \end{aligned}
\end{equation}
\begin{figure}
 \begin{center}
  \includegraphics[width=.7\linewidth]{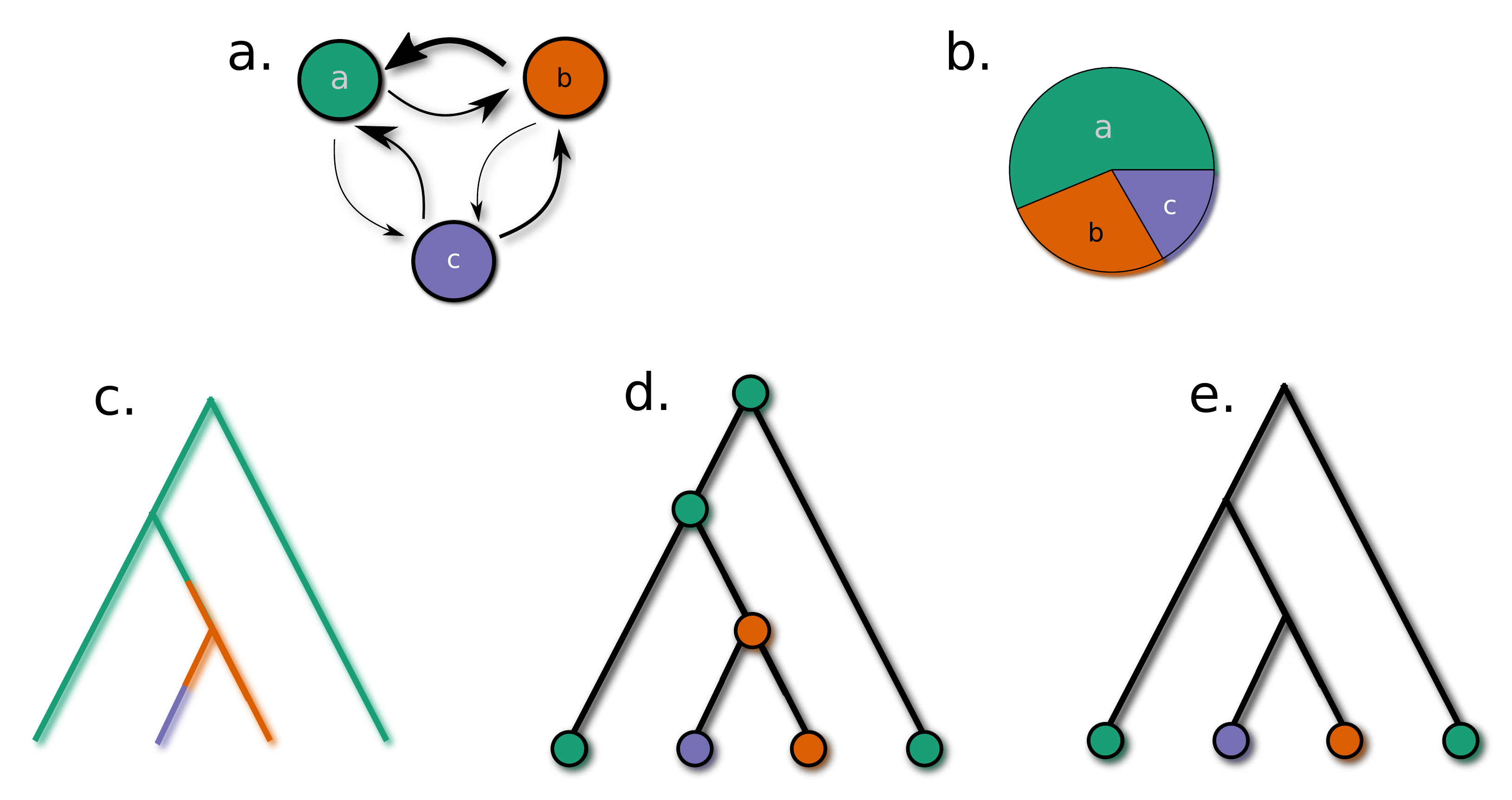}
 \end{center}
 \caption{ a. CTMC b. Equilibrium distribution c. Fully specified history of a
  phylogenetic Markov chain d. Marginalizing over events at branches e. Marginalizing
  over states at internal nodes}\label{fig:2}
\end{figure}
The transition rates and the equilibrium distribution are illustrated in the upper panels
of Figure \ref{fig:2}.

A complete history of a run of this CTMC along the branches of a phylogeny is shown in
panel c.\ of Figure \ref{fig:2}. If the transition rates and branch lengths are known, the
likelihood of this history, conditional on the state at the root, can be computed. To
completely specify the likelihood of the history, one needs the probability distribution
over states at the root of the tree --- i.e., at the proto-language. In this paper we
assume that proto-languages of known language families are the result of a long time of
language change. If nothing about this history is known, the distribution of states at the
proto-language is therefore virtually identical to the equilibrium
distribution.\footnote{It is possible to test for a given feature distribution and
 phylogeny whether the data support this hypothesis. We leave this issue for future
 work.}

The precise points in time where state transitions occur are usually unknown though. We
can specify an infinite set of possible histories which only agree on the states of the
nodes of the tree (illustrated in panel d.\ of Figure \ref{fig:2}). The marginal
likelihood of this set is the product of the conditional likelihood of the bottom node of
each branch given the top node and the length of each branch, multiplied with the
equilibrium probability of the root state.

Under normal circumstances only the states of the extant languages, i.e.\ at the leaves of
the tree, are known. The marginal likelihood of all histories agreeing only in the states
at the leaves can be determined by summing over all possible distribution of states at
internal nodes (illustrated in panel e.\ of Figure \ref{fig:2}). This quantity can be
computed efficiently via a recursive bottom-up procedure known as
\citeauthor{felsenstein81}'s (\citeyear{felsenstein81}) \emph{pruning algorithm}.

\begin{figure}
 \begin{center}
  \includegraphics[width=7cm]{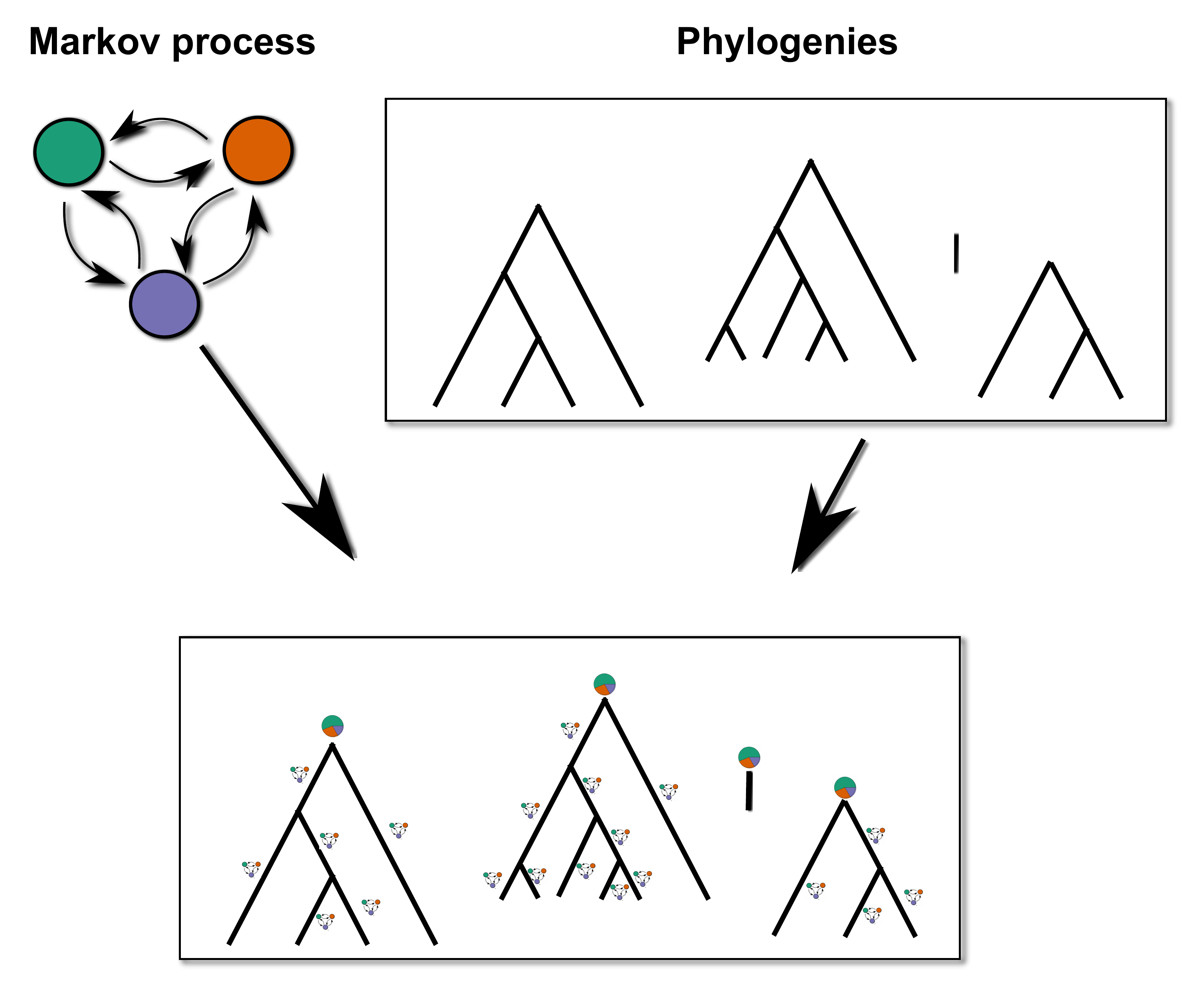}
 \end{center}
 \caption{ Phylogenetic Markov CTMC with a collection of phylogenies}\label{fig:3}
\end{figure}

This can easily be extended to sets of phylogenies (e.g., a collection of phylogenies for
different language families; schematically illustrated in Figure \ref{fig:3}
).
Language isolates are degenerate phylogenies with only one leave that is also the
root. The likelihood of the state of an isolate is thus the equilibrium probability of its
state.

Under the assumption that the distributions in different language families are
independent, the total likelihood of such a collection of phylogenies is the product of
the individual tree likelihoods.

Under realistic conditions, the precise phylogeny of a language family is never
known. Rather, it is possible to infer a probability distribution over phylogenies using
Bayesian inference and, e.g., lexical data. In such a scenario, the \emph{expected
 likelihood} for a language family is the averaged likelihood over this distribution of
trees.

If only the phylogeny and the states at the leaves are known, statistical inference can be
used to determine the transition rates (and thereby also the equilibrium
distribution). Bayesian inference, that is used in this study, requires to specify prior
assumptions over the transition rates and results in a posterior distribution over these
rates.

In the remainder of this paper, this program is illustrated with a case on word order
features and their potential correlations.\\

\subsection{Word order features}
\label{subsec:2_3}

The typical order of major syntactic constituents in declarative sentences of a language,
and the distribution of these properties across the languages of the world, has occupied
the attention of typologists continuously since the work of \cite{greenberg63} (see, e.g.,
\citealt{lehmann73,vennemann74,hawkins83,dryer92}, among many others). There is a
widespread consensus that certain word-order features are typologically correlated. For
instance, languages with verb-object order tend to be prepositional while object-verb
languages are predominantly postpositional. Other putative correlations, like the one
between verb-object order and adjective-noun order, are more controversial.

The study in \cite{dunnetal11} undermined this entire research program. They considered
eight word-order features and four major language families. For each pair of features and
each family, they conducted a statistical test whether the feature pair is correlated in
that family, using Bayesian phylogenetic inference. Surprisingly, they found virtually no
agreement across language families. From this finding they conclude that the dynamics of
change of word-order features is lineage specific; so the search for universals is void.

We will take up this problem and will consider the same eight word order features, which
are taken from the \emph{World Atlas of Language Structures} (WALS; \citealt{wals13}). For
each of the 28 feature pairs, we will test two hypotheses:

\begin{enumerate}
 \item All lineages (language families and isolates) share the parameters of a CTMC
       governing the evolution of these features (vs.\ Each lineage has its own CTMC
       parameters), and
 \item If all lineages share CTMC parameters, the two features are correlated.
\end{enumerate}

For each of the eight features considered, only the values ``head-dependent'' vs.\
``dependent-head'' are considered. Languages that do not fall in either category are
treated as ``missing value''. These features and the corresponding values are listed in
Table \ref{tab:1}.\\

\begin{table}
 \begin{center}
  \begin{tabular}{lll}
   \toprule{}
   \emph{feature} & \emph{value 1}       & \emph{value 2}       \\\midrule
   VS             & verb-subject         & subject-verb         \\
   VO             & verb-object          & object-verb          \\
   PN             & adposition-noun      & noun-adposition      \\
   NG             & noun-genitive        & genitive-noun        \\
   NA             & noun-adjective       & adjective-noun       \\
   ND             & noun-demonstrative   & demonstrative-noun   \\
   NNum           & noun-numeral         & numeral-noun         \\
   NRc            & noun-relative clause & relative clause-noun \\\bottomrule
  \end{tabular}
 \end{center}
 \caption{Word order features}\label{tab:1}
\end{table}

\subsection{Obtaining language phylogenies}
\label{subsec:2_4}
In \citep{jaeger18scientificData} a method is described how to extract binary characters
out of the lexical data from the \emph{Automated Similarity Judgment Program} (ASJP v.\
18; \citealt{asjp18}). These characters are suitable to use for Bayesian phylogenetic
inference.

The processing pipeline described in \citep{jaeger18scientificData} is briefly recapitulated
here. The ASJP data contains word lists from more than 7,000 languages and dialects,
covering the translations of 40 core concepts. All entries are given in a uniform phonetic
transcription.

In a first step, mutual string similarities are computed using pairwise sequence alignment
along the lines of \citep{jaeger13ldc}. From these similarities, pairwise language
distances are computed. These two measures are used to group the words for a each concept
into cluster approximating \emph{cognate classes}. Each such cluster defines a binary
character, with value 1 for languages containing an element of the cluster in its word
list, 0 if the entries for the same concepts all belong to different clusters, and
undefined if there is no entry for that concept.

An additional class of binary characters is obtained from the Cartesian product of the 40
concepts and the 41 sound classes used in ASJP. A language has entry 1 for character
``concept c/sound class s'' if one of the entries for concept ``c'' contains at least one
occurrence of sound class ``s'', 0 if none of the entries for ``c'' contain ``s'', and
undefined if there is no entry for that concept.

In \citep{jaeger18scientificData} it is demonstrated that phylogenetic inference based on
these characters is quite precise. For this assessment, the expert classification from \citep{glottolog3_3_2} is used as gold standard.

\begin{figure}
 \begin{center}
  \includegraphics[width=\linewidth]{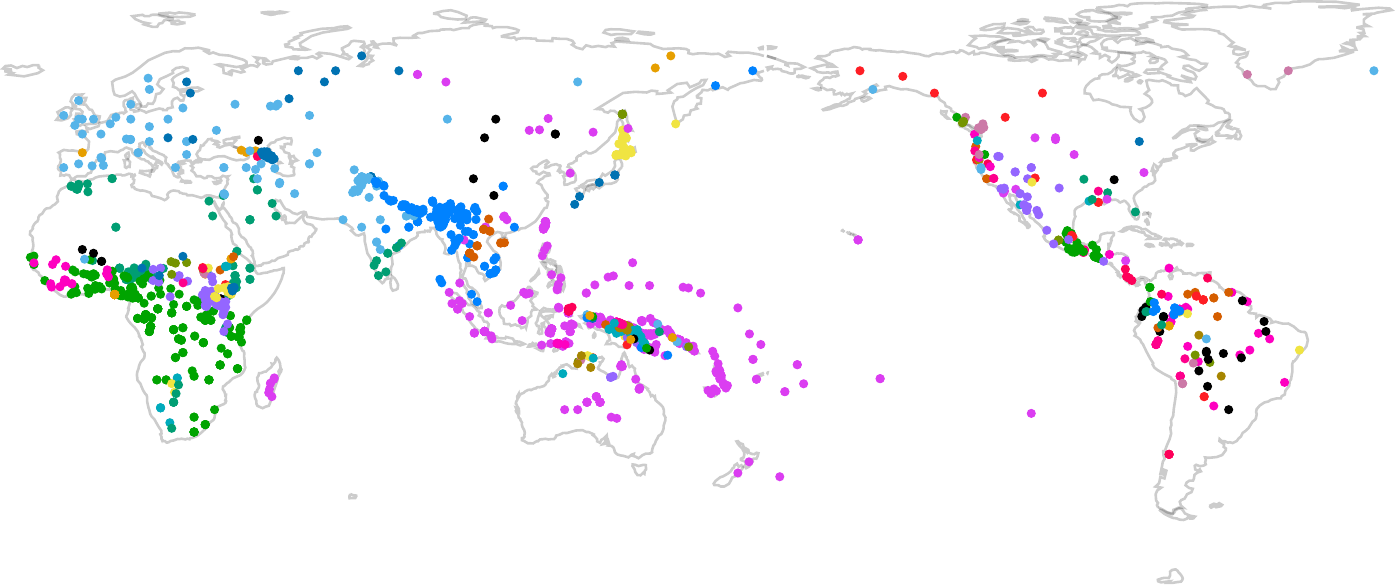}
 \end{center}
 \caption{ Geographic distribution of the sample of languages used. Colors indicate Glottolog classification}\label{fig:4}
\end{figure}

For the present study, we identified a total of 1,626 of languages for which WALS contains information about at least one word-order feature and the data from \citep{jaeger18scientificData} contain characters. These languages comprise 175 lineages according to the Glottolog classification, including 81 isolates.\footnote{A language is called an isolate here if our 1,626-language sample contains no other language belonging to the same Glottolog family.} The geographic distribution of this sample is shown in Figure \ref{fig:4}.

For the present study, we used the cognate classes occurring within the language sample, as well as the concept/sound class characters as input for Bayesian phylogenetic inference. For each language family, a posterior tree sample was inferred using the Glottolog classification as constraint trees.\footnote{For this purpose we used the software \emph{MrBayes} \citep{mrbayes3} for families with at least three members and \emph{RevBayes} \citep{revbayes} for two-language families.} For each family, we sampled 1,000 phylogenies from the posterior distribution for further processing.\\

\subsection{Generative models}

\begin{figure}
 \begin{center}
  \includegraphics[width=6cm]{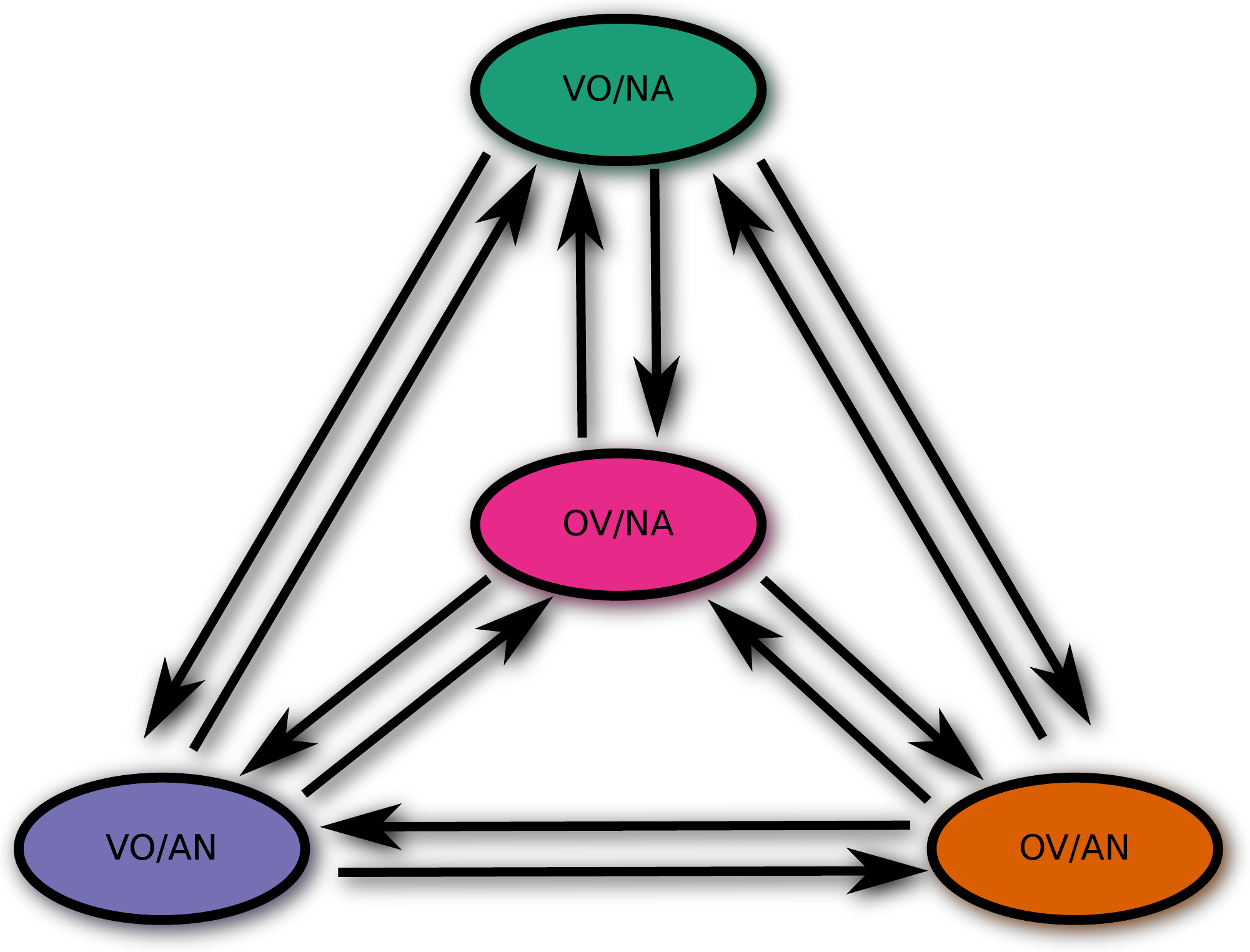}
 \end{center}
 \caption{ CTMC for a possibly correlated feature pair}\label{fig:5}
\end{figure}

To study the co-evolution of two potentially correlated word-order features, we assume a four-state CTMC for each pair of such features --- one state for each combination of values. We assume that all twelve state transitions are \emph{a priori} possible, including simultaneous changes of both feature values.\footnote{\cite{dunnetal11}, following the general methodology of \cite{pagelMeade06}, exclude this possibility. We believe, however, that this possibility should not be excluded \emph{a priori} because it is conceivable that during a major reorganization of the grammar of a language, several features change their values at once.}%
The structure of the CTMC is schematically shown in Figure \ref{fig:5} for the feature pair VO/NA.

As pointed out above, \cite{dunnetal11} argue that the transition rates between the states of word-order features follow lineage-specific dynamics. To test this assumption (Hypothesis 1 above), we fitted two models for each feature pair:
\begin{itemize}
 \item a \textbf{universal model} where all lineages follow the same CTMC with universally identical transition rates, and
 \item a \textbf{lineage-specific model} where each lineage has its own set of transition rate parameters.
\end{itemize}

\begin{figure}
 \begin{center}
  \includegraphics[width=\linewidth]{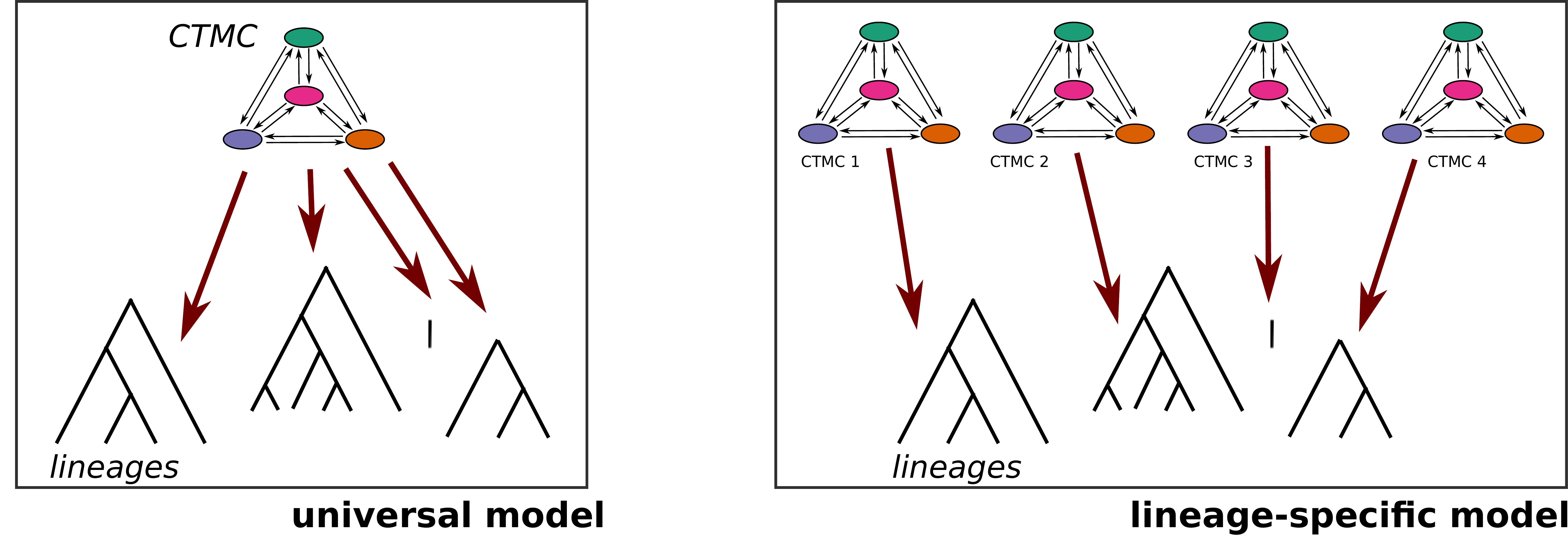}
 \end{center}
 \caption{ Universal vs.\ lineage-specific model}\label{fig:6}
\end{figure}

These two model structures are illustrated in Figure \ref{fig:6}.

For all models we chose a log-normal distribution with parameters $\mu=0$ and $\sigma=1$ as prior for all rate parameters.

We will determine via statistical model comparison for each feature pair which of the two models fits the data better.\\

\subsection{Prior predictive sampling}

\label{subsec:2_6}

In a first step, we performed prior predictive sampling for both model types. This means that we simulated artificial datasets that were drawn from the prior distributions, and then compared them along several dimensions with the empirical data. This step is a useful heuristics to assess whether the chosen model and the chosen prior distributions are in principle capable to adequately model the data under investigation.

We identified three statistics to summarize the properties of these artificial data and compare them with the empirically observed data. For this purpose we represented each language as a probability vector over the four possible state. Let $Y$ be the data matrix with languages as rows and states as columns, and $n$ the number of languages, and $F$ the set of lineages, where each lineage is a set of languages. The statistics used are:
\begin{itemize}
 \item the \textbf{total variance}:
       $$
        \frac{1}{n}\sum_i (\sum_l Y_{l,i}^2 - (\sum_l Y_{l,i})^2)
       $$
 \item the \textbf{mean lineage-wise variance}:
       $$
        \frac{1}{|F|}\sum_{k}\frac{1}{|F_k|}\sum_i (\sum_{l\in F_k} Y_{l,i}^2 - (\sum_{l\in F_k} Y_{l,i})^2)
       $$
 \item the \textbf{cross-family variance}, i.e. the total variance between the centroids for each lineage:
       $$
        \sum_i(\frac{1}{|F|}\sum_k(\frac{1}{|F_k|} \sum_{l\in F_k} Y_{l,i})^2 -
        (\frac{1}{|F|}\sum_k\frac{1}{|F_k|} \sum_{l\in F_k} Y_{l,i})^2)
       $$
\end{itemize}

\begin{figure}
 \begin{center}
  \includegraphics[width=\linewidth]{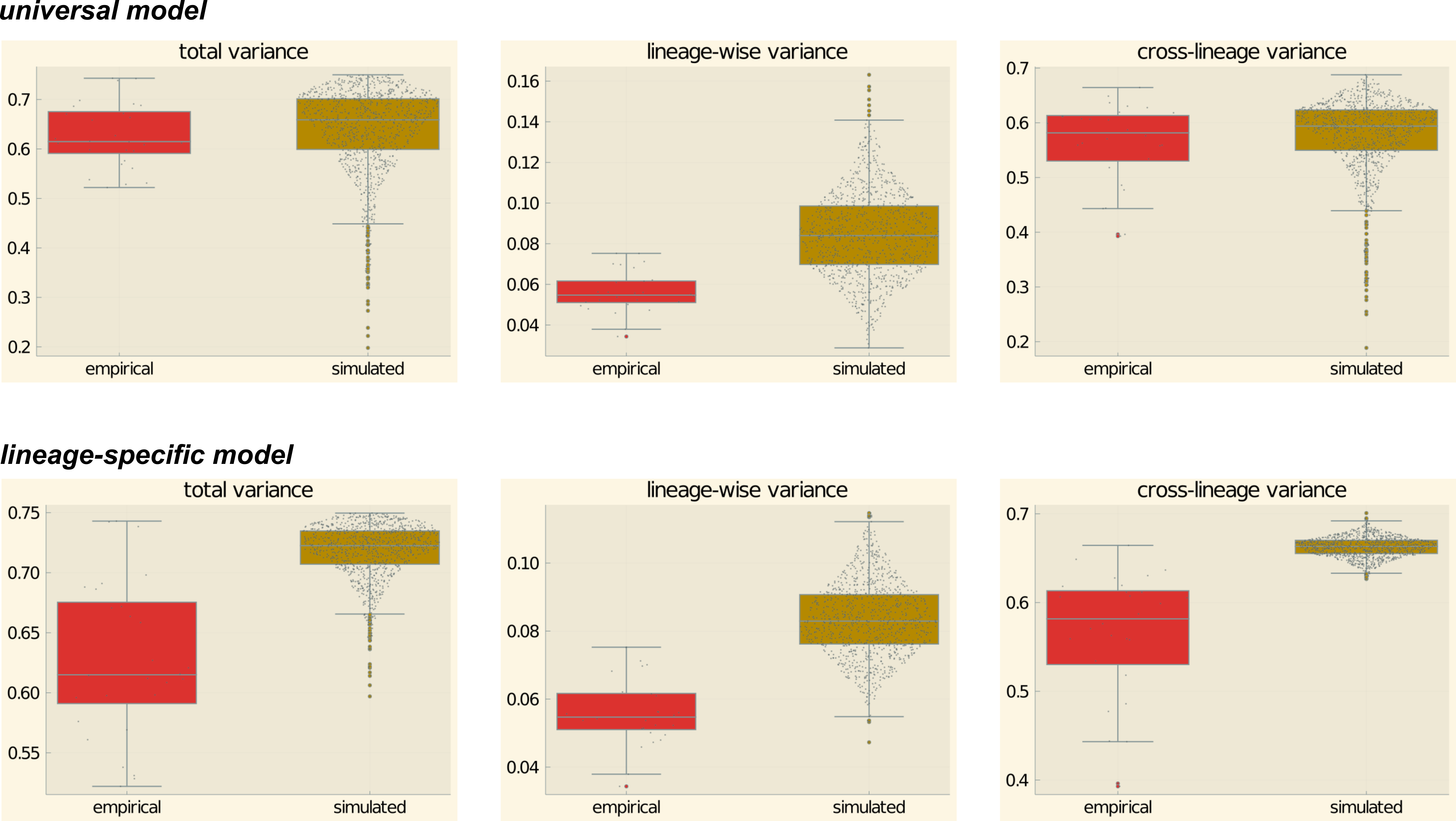}
 \end{center}
 \caption{ Prior predictive simulations}\label{fig:7}
\end{figure}

In Figure \ref{fig:7}, the distribution of these statistics for the 28 feature pairs for the empirical data are compared with the prior distributions for the universal model (top panels) and the lineage-specific model (bottom panels). For each model, we conducted 1,000 simulation runs.

From visual inspection it is easy to see that for the universal model, the empirically observed values fall squarely within the range of the prior distributions. For the lineage-specific model, the observed variances are generally lower than than expected under the prior distribution. This is especially obvious with regard to the cross-family variance, which is much lower for the empirical data than predicted by the model.

\subsection{Model fitting}

Both models where fitted for each of the 28 feature pairs. Computations were performed using the programming language \emph{Julia} and Brian J.\ Smith's package \emph{Mamba} (\url{https://github.com/brian-j-smith/Mamba.jl}) for Bayesian inference. We extended \emph{Mamba} by functionality to handle phylogenetic CTMC models.

Posterior samples were obtained via Slice sampling \citep{neal2003}. Averaging over the prior of phylogenies was achieved by randomly picking one phylogeny from the prior (see Subsection \ref{subsec:2_4}) in each MCMC step. Posterior sampling was stopped when the \emph{potential scale reduction factor} (PSRF; \citealt{gelmanrubin92}) was $\leq 1.1$ for all parameters.\\

\subsection{Posterior predictive sampling}

To test the fit of the models to the data, we performed \emph{posterior predictive sampling} for all fitted models. This means that for each model, we randomly picked 1,000 samples from the posterior distribution and used it to simulate artificial datasets. The three statistics used above for prior predictive sampling were computed for each simulation. The results are shown in
the Supplementary Material.

With regard to total variance, we find that the empirical value falls outside the $95\%$ highest posterior interval for three out of 28 feature pairs (VO-NRc, PN-NRc and NA-ND), where the model overestimates the total variance. The lineage-specific model overestimates the total variance for ten feature pairs.

Since three outliers out of 28 is within the expected range for a $95\%$ interval, we can conclude that the universal model generally predicts the right amount of cross-linguistic variance. The lineage-specific model overestimates this quantity.

For cross-linguistic variance, the empirical value falls outside the HPD ($95\%$ highest posterior density interval) for 14 pairs for the universal model and for 21 pairs for the lineage-specific model. So both models tend to overestimate this variable. This might be due to the fact that phylogenetic CTMC models disregard the effect of language contact, which arguably reduced within-family variance.

The cross-family variance falls into the universal model's HPD for all pairs, but only for two pairs (VO-NA, VO-NNum) for the lineage-specific model. Briefly put, the universal model gets this quantity right while the lineage-specific model massively overestimates it.\\

\subsection{Bayesian model comparison}
\label{subsec:2_9}

As a next step we performed statistical model comparison between the universal and the lineage-specific model. Briefly put, model comparison estimates how well models will serve to predict unseen data that are generated by the same process as the observed data, and compares the predictive performances. Everything else being equal, the model with the better predictive performance can be considered a better explanation for the observed data.

Since there is no general consensus about the best method to compare Bayesian models (see, e.g., \citep{vehtariOjanen2012} for an overview), we applied two different techniques.

The \emph{marginal likelihood} of the data under a Bayesian model is the expected likelihood of the data $y$ weighted by the prior probability of the model parameters $\theta$.

$$
 D(y|M) = \int_{\theta}p(y|\theta)p(\theta|M)d\theta
$$

The Bayes factor between two models $M_1$ and $M_2$ is the ratio of their marginal densities:
$$
 BF = \frac{D(y|M_1)}{D(y|M_2)}
$$

To estimate the marginal densities, we used \textbf{bridge sampling} (cf.\ \citealt{gronauetal2017}). For our implementation we depended strongly on the \emph{R}-package \emph{bridgesampling} \citep{bridgesampling}. The logarithmically transformed Bayes factors between the universal model ($\approx M_1$) and the lineage-specific model ($\approx M_2$) are shown for each feature pair in Table \ref{tab:5}.

\begin{table}
 \begin{center}
  \begin{tabular}{lrl}
   \toprule
   \em feature pair & \em (log) Bayes Factor & \em cumulative posterior probability \\\midrule
   VS-VO            & 72.9                   & 0.000                                \\
   VS-NG            & 65.9                   & 0.000                                \\
   PN-NG            & 64.5                   & 0.000                                \\
   VO-PN            & 56.8                   & 0.000                                \\
   VS-PN            & 54.4                   & 0.000                                \\
   VO-NG            & 41.3                   & 0.000                                \\
   VS-NRc           & 36.5                   & 1.11e-16                             \\
   NA-ND            & 32.1                   & 1.18e-14                             \\
   VS-NNum          & 31.1                   & 4.19e-14                             \\
   VS-NA            & 30.8                   & 8.57e-14                             \\
   NG-NRc           & 28.0                   & 8.09e-13                             \\
   VO-NRc           & 27.7                   & 1.79e-12                             \\
   VS-ND            & 27.0                   & 3.67e-12                             \\
   PN-NRc           & 25.6                   & 1.12e-11                             \\
   NA-NRc           & 22.1                   & 2.63e-10                             \\
   NG-ND            & 19.0                   & 5.98e-9                              \\
   NG-NA            & 18.8                   & 1.29e-8                              \\
   ND-NNum          & 15.6                   & 1.76e-7                              \\
   PN-NA            & 15.2                   & 4.38e-7                              \\
   NA-NNum          & 8.8                    & 0.000147                             \\
   PN-ND            & 8.7                    & 0.000319                             \\
   ND-NRc           & 7.3                    & 0.00101                              \\
   NG-NNum          & 6.7                    & 0.00223                              \\
   VO-ND            & 6.4                    & 0.00393                              \\
   NNum-NRc         & 5.3                    & 0.00892                              \\
   PN-NNum          & 5.1                    & 0.0152                               \\
   VO-NA            & 5.0                    & 0.0218                               \\
   VO-NNum          & 2.4                    & 0.102                                \\\bottomrule
  \end{tabular}
 \end{center}
 \caption{log-Bayes factor between universal and lineage-specific model. The last row gives the upper limit of the posterior probability that for at least one feature-pair up to this line the lineage-specific model is true.}\label{tab:5}
\end{table}
%

All log-Bayes factors are positive, i.e., favor the universal over the lineage-specific model.

According to the widely used criteria by \cite{jeffrey98}, a Bayes factor of $\geq 100$, which corresponds to a logarithmic Bayes factor of $4.6$, is considered as decisive evidence. So except for the feature pair VO-NNum, this test provides decisive evidence in favor of the universal model.

Unlike frequentist hypothesis testing, Bayesian model comparison does not require a correction for multiple testing. Still, since 28 different hypotheses are tested simultaneously here, the question arises how confident we can be that a given subset of the hypotheses are true. Assuming the uninformative prior that the universal and the lineage-specific model are equally likely \emph{a priori}, the posterior probability of the universal model being true given that one of the two models is true, is the logistic transformation of the log-Bayes factor. Let us call this quantity $p^u_i$ for feature pair $i$. We assume that feature-pairs are sorted in descending order according to their Bayes factor, as in Table \ref{tab:5}. The posterior probability of the lineage-specific model is $p^l_i = 1-p^u_i$. The quantity $p^l_{1\cdots k}$ is the cumulative probability that the lineage-specific model is true for at least one feature pair $i$ with $1\leq i\leq k$.\footnote{This amounts to the Holm-Bonferroni correction \citep{holm1979}, but we use it here to compute an upper limit for the posterior probability rather than for the expected $\alpha$ level.}

Since the hypotheses for the individual feature pairs are not mutually independent, it is not possible to compute this probability, However, according to the \emph{Bonferroni inequality},
$$
 p^l_{1\cdots k} \leq \sum_{1\leq i\leq k}p^l_k
$$
This upper limit is shown in the third column of Table \ref{tab:5}. For all but the feature pair VO-NNum, this probability is $< 0.05$. We conclude that this line of reasoning also confirms that the data strongly support the universal over the lineage-specific hypothesis for all feature pairs except VO-NNum.

As alternative approach to model comparison, we conducted \textbf{Pareto-smoothed cross-validation} \citep{vehtarietal2017} using the \emph{R}-package \emph{loo} \citep{loo}.

Leave-one-out cross-validation means to loop through all data points $y_i$ and compute the quantity
$$
 \log p(y_i|y_{-i}) = \int_\theta p(y_i|\theta)p(\theta|y_{-i})d\theta
$$

Here, $y_{-i}$ denotes the collection of all datapoints $\neq y_i$. Since this amounts to fitting a posterior distribution as often as there are datapoints, this is computationally not feasible in most cases (including the present case study). The quantity
$$
 \mathrm{elpd} = \sum_i \log p(y_i|y_{-i}),
$$
the \emph{expected log pointwise predictive density}, is a good measure of how well a model predicts unseen data and can be used to compare models.

Since computing the elpd amounts to fitting a posterior distribution for each datapoint, the method is not feasible though in most cases (including the present case study). Pareto-smoothed leave-one-out cross-validation is a technique to estimate elpd from the posterior distribution of the entire dataset.

However, this algorithm depends on the assumption that individual datapoints are mutually \emph{conditionally independent}, i.e.,
$$
 p(y|\theta) = \prod_i p(y_i|\theta).
$$

This is evidently not the case for phylogenetic CTMC models if we treat each language as a datapoint.\footnote{To see why, consider an extreme case where the phylogeny consists of two leaves with an infinitesimally small co-phenetic distance, and the equilibrium distribution over states is the uniform distribution. Then $p(y_1=y_2=s_1) \approx p(y_1=s_1) = p(y_2=s_1) < p(y_1=s_1)p(y_2=s_1)$.} However, conditional independence does hold between lineages both in the universal and the lineage-dependent model. Pareto-smoothed leave-one-out cross-validation can be therefore be performed if entire lineages are treated as datapoints.

The difference in elpd, i.e., elpd of universal model minus elpd of lineage-specific model, are shown in Table \ref{tab:6}.
\begin{table}
 \begin{center}
  \begin{tabular}{lr}
   \toprule
   \em feature pair & \em $\Delta$ elpd \\\midrule
   VS-VO            & 79.7              \\
   PN-NG            & 75.9              \\
   VS-NG            & 72.6              \\
   VO-PN            & 65.3              \\
   VS-PN            & 61.4              \\
   VO-NG            & 48.3              \\
   VS-NRc           & 45.5              \\
   NA-ND            & 37.3              \\
   NG-NRc           & 36.7              \\
   VS-NA            & 35.3              \\
   VO-NRc           & 34.6              \\
   PN-NRc           & 32.7              \\
   VS-NNum          & 28.7              \\
   NA-NRc           & 27.9              \\
   VS-ND            & 26.1              \\
   NG-NA            & 21.1              \\
   PN-ND            & 19.2              \\
   PN-NA            & 18.0              \\
   NG-ND            & 16.1              \\
   VO-ND            & 13.6              \\
   ND-NNum          & 12.5              \\
   NG-NNum          & 12.3              \\
   ND-NRc           & 7.5               \\
   PN-NNum          & 6.6               \\
   NA-NNum          & 4.2               \\
   NNum-NRc         & 3.7               \\
   VO-NNum          & 3.5               \\
   VO-NA            & 1.1               \\\bottomrule
  \end{tabular}
 \end{center}
 \caption{Differences in elpd}\label{tab:6}
\end{table}
%
For all feature pairs, the elpd is higher for the universal than for the lineage-specific model.\\

\subsection{Feature correlations}
\label{subsec:2_10}

Let us know turn to the second hypothesis mentioned in Subsection \ref{subsec:2_3}, repeated here. For each feature pair, we will probe the question:
\begin{enumerate}
 \item[2.] If all lineages share CTMC parameters, the two features are correlated.
\end{enumerate}

\begin{figure}
 \begin{center}
  \includegraphics[width=.7\linewidth]{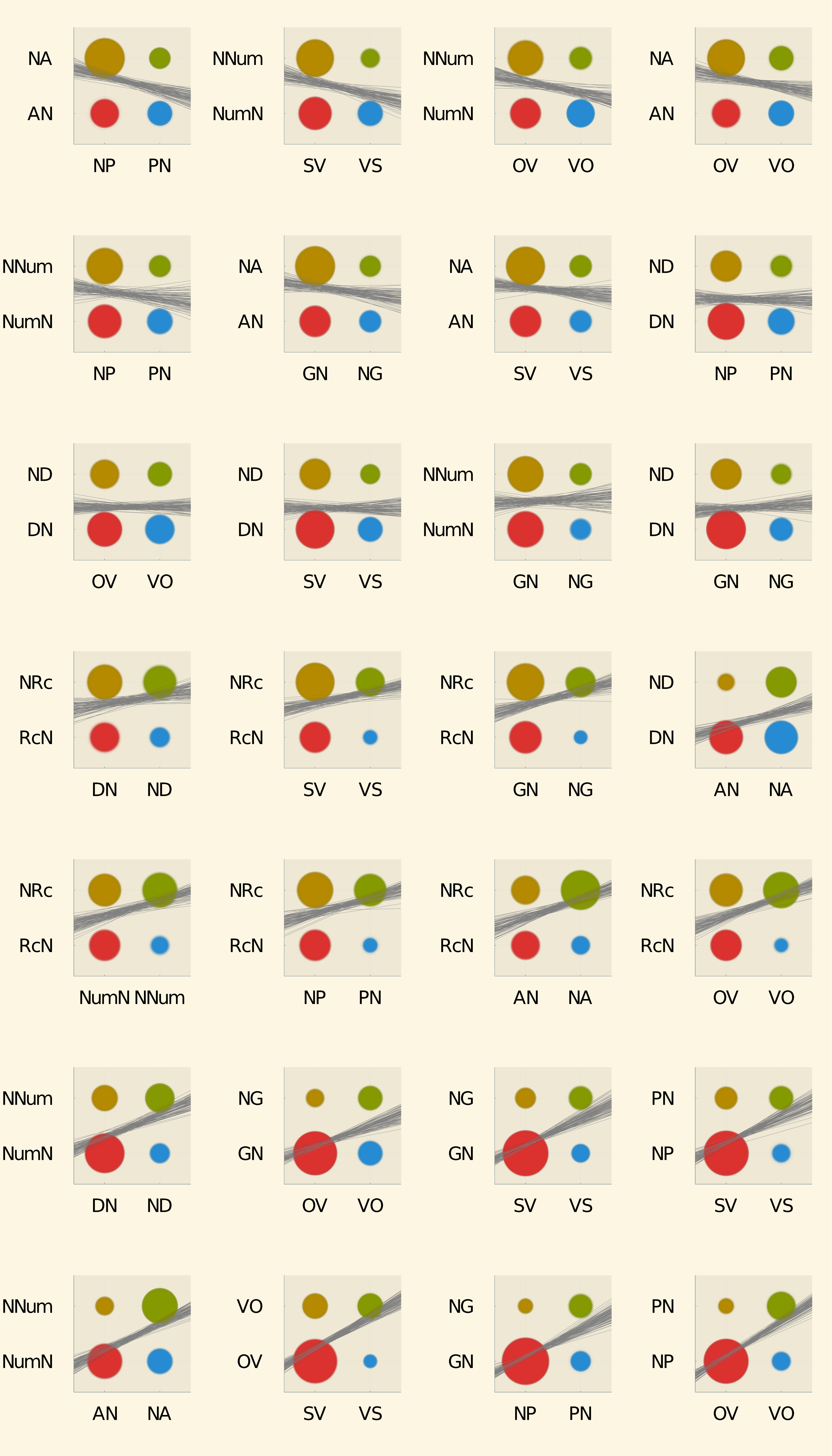}
 \end{center}
 \caption{ Posterior equilibrium probabilities and linear regression}\label{fig:13}
\end{figure}

To operationalize correlation, we define the feature value ``dependent precedes head'' as 0 and ``head precedes dependent'' as 1. For a given feature pair, this defines a $2\times 2$ contingency table with posterior equilibrium probabilities for each value combination. They are displayed in Figure \ref{fig:13}. In each diagram, the $x$-axis represents the first feature and the $y§$-axis the second feature. The size of the circles at the corners of the unit square indicate the equilibrium probability of the corresponding value combination. Blurred edges of the circles represent posterior uncertainty.
The diagrams also show the posterior distribution of regression lines indicating the direction and strength of the association between the two features.\footnote{%
Intercept and slope of the regression lines are $\frac{p_{01}}{p_{00}+p_{01}}$ and $\frac{p_{11}}{p_{10}+p_{11}} - \frac{p_{01}}{p_{00}+p_{01}}$ respectively.} Perhaps surprisingly, for some feature pairs the association is negative.

The \emph{correlation} between two features binary $f_1, f_2$ in the strict mathematical sense, also called the \emph{Phi coefficient}, is
$$
 \frac{\mathrm{cov}(f_1,f_2)}{\sqrt{\mathrm{var}(f_1)\mathrm{var}(f_2)}} = \frac{p_{00}p_{11} - p_{10}p_{01}}{\sqrt{(p_{00}+p_{01})(p_{10}+p_{11})(p_{00}+p_{10})(p_{10}+p_{11})}}
$$
and ranges from $-1$ (perfect negative relationship) to $1$ (perfect positive relationship), with $0$ indicating no relationship.

\begin{figure}
 \begin{center}
  \includegraphics[width=8cm]{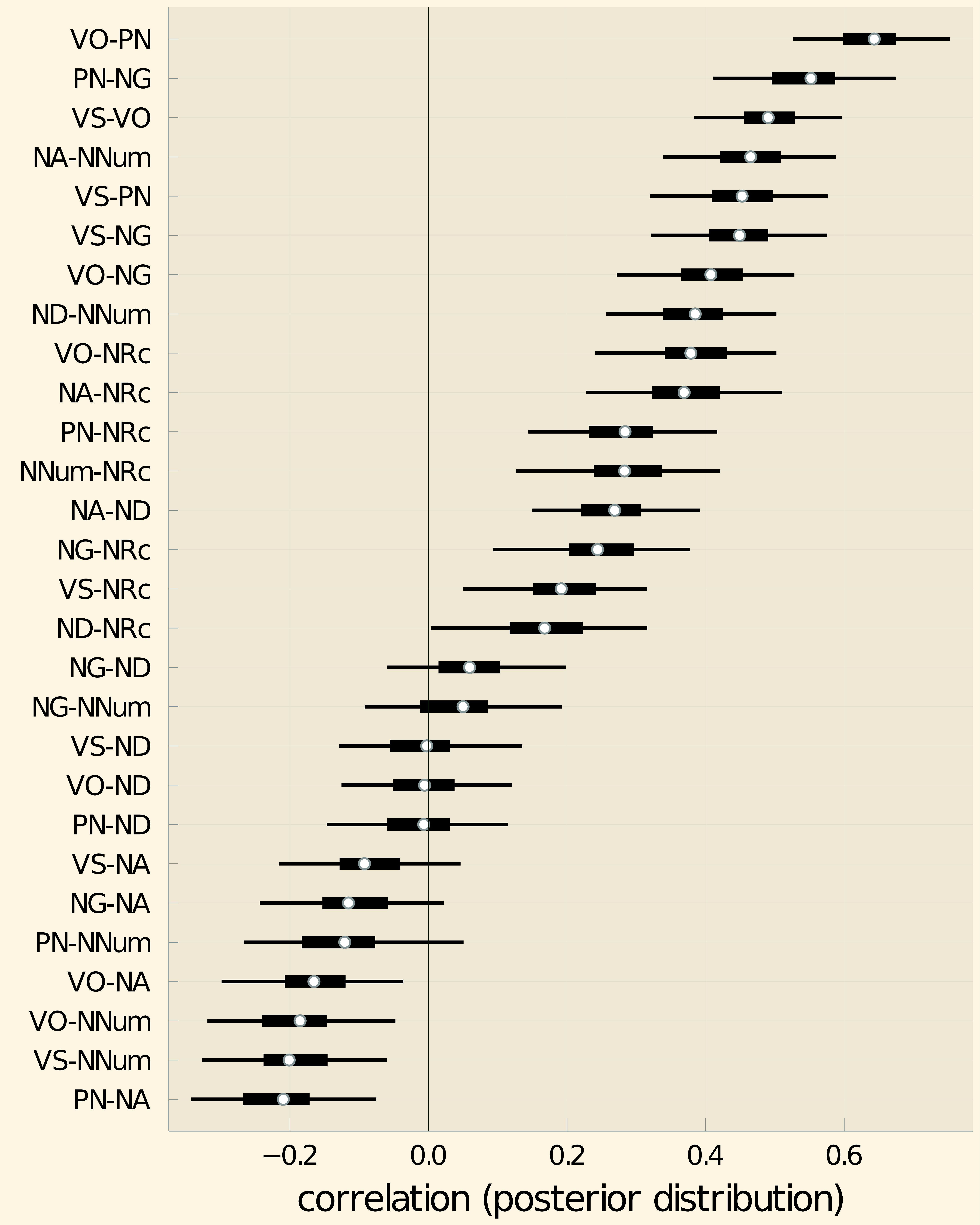}
 \end{center}
 \caption{ Correlation coefficients for feature pairs. White dots indicate the median, thick lines the $50\%$ and thin lines the $95\%$ HPD intervals.}\label{fig:14}
\end{figure}

The median posterior correlations and the corresponding HPD interval given in Table \ref{tab:7} and shown in Figure \ref{fig:14}.

\begin{table}
 \begin{center}
  \begin{tabular}{lrr}
   \toprule
   \em feature pair & \em median & HPD            \\\midrule
   VO-PN            & 0.64       & (0.53, 0.75)   \\
   PN-NG            & 0.55       & (0.41, 0.67)   \\
   VS-VO            & 0.49       & (0.38, 0.60)   \\
   NA-NNum          & 0.47       & (0.34, 0.59)   \\
   VS-PN            & 0.45       & (0.32, 0.58)   \\
   VS-NG            & 0.45       & (0.32, 0.58)   \\
   VO-NG            & 0.41       & (0.27, 0.53)   \\
   ND-NNum          & 0.38       & (0.26, 0.50)   \\
   VO-NRc           & 0.38       & (0.24, 0.50)   \\
   NA-NRc           & 0.37       & (0.23, 0.51)   \\
   PN-NRc           & 0.28       & (0.14, 0.42)   \\
   NNum-NRc         & 0.28       & (0.13, 0.42)   \\
   NA-ND            & 0.27       & (0.15, 0.39)   \\
   NG-NRc           & 0.24       & (0.09, 0.38)   \\
   VS-NRc           & 0.19       & (0.05, 0.32)   \\
   ND-NRc           & 0.17       & (0.00, 0.32)   \\
   NG-ND            & 0.06       & (-0.06, 0.20)  \\
   NG-NNum          & 0.05       & (-0.09, 0.19)  \\
   VS-ND            & -0.00      & (-0.13, 0.14)  \\
   VO-ND            & -0.01      & (-0.13, 0.12)  \\
   PN-ND            & -0.01      & (-0.15, 0.11)  \\
   VS-NA            & -0.09      & (-0.22, 0.05)  \\
   NG-NA            & -0.12      & (-0.24, 0.02)  \\
   PN-NNum          & -0.12      & (-0.27, 0.05)  \\
   VO-NA            & -0.17      & (-0.30, -0.04) \\
   VO-NNum          & -0.19      & (-0.32, -0.05) \\
   VS-NNum          & -0.20      & (-0.33, -0.06) \\
   PN-NA            & -0.21      & (-0.34, -0.08) \\\bottomrule
  \end{tabular}
 \end{center}
 \caption{ Correlation coefficients for feature pairs: Median and $95\%$ HPD interval}\label{tab:7}
\end{table}

How reliable are these estimates? The Bayes factor between the hypotheses ``correlation $\neq 0$'' and ``correlation $=0$'' can be determined via the Savage-Dickey method (\citealt{dickey70}; see also \citealt{wagenmakeretal2020}). We used the \emph{R}-package \emph{LRO.utilities} (\url{https://github.com/LudvigOlsen/LRO.utilities/}) to carry out the computations. The log-Bayes factors for the individual feature pairs are shown in Table \ref{tab:8}.

\begin{table}
 \begin{center}
  \begin{tabular}{lrl}
   \toprule
   \em feature pair & (log) Bayes Factor & \em cumulative posterior probability \\\midrule
   VO-PN            & 19.25              & \textbf{4.37e-9}                     \\
   VS-VO            & 15.82              & \textbf{1.39e-7}                     \\
   VS-NG            & 14.12              & \textbf{8.78e-7}                     \\
   PN-NG            & 12.07              & \textbf{6.62e-6}                     \\
   VS-PN            & 12.03              & \textbf{1.26e-5}                     \\
   NA-NNum          & 10.93              & \textbf{3.05e-5}                     \\
   ND-NNum          & 9.98               & \textbf{7.68e-5}                     \\
   VO-NG            & 8.85               & \textbf{0.00022}                     \\
   VO-NRc           & 8.15               & \textbf{0.000509}                    \\
   NA-NRc           & 7.43               & \textbf{0.0011}                      \\
   NA-ND            & 4.72               & \textbf{0.00995}                     \\
   NNum-NRc         & 4.07               & \textbf{0.0267}                      \\
   PN-NRc           & 3.83               & \textbf{0.0479}                      \\
   PN-NA            & 3.17               & 0.0884                               \\
   NG-NRc           & 2.64               & 0.155                                \\
   VS-NNum          & 2.47               & 0.233                                \\
   VO-NNum          & 2.01               & 0.351                                \\
   VS-NRc           & 1.93               & 0.478                                \\
   VO-NA            & 1.60               & 0.646                                \\
   ND-NRc           & 0.55               & 1.000                                \\
   NG-NA            & -0.03              & 1.000                                \\
   PN-NNum          & -0.43              & 1.000                                \\
   VS-NA            & -0.51              & 1.000                                \\
   NG-ND            & -1.17              & 1.000                                \\
   NG-NNum          & -1.32              & 1.000                                \\
   PN-ND            & -1.52              & 1.000                                \\
   VO-ND            & -1.56              & 1.000                                \\
   VS-ND            & -1.64              & 1.000                                \\\bottomrule
  \end{tabular}
 \end{center}
 \caption{log-Bayes factor between ``correlation $\neq 0$'' and ``correlation $=0$''. The last row gives the upper limit of the posterior probability that for at least one feature-pair up to this line correlation $=0$.}\label{tab:8}
\end{table}
%

Using the same method as in Subsection \ref{subsec:2_9}, we can conclude with $95\%$ confidence that there is a non-zero correlation for 13 feature pairs: VO-PN, VS-VO, VS-NG, PN-NG, NA-NNum, ND-NNum, VO-NG, VO-NRc, NA-NRc, NA-ND, NNum-NRc, PN-NRc. For all these pairs, the correlation coefficient is credibly positive (meaning the $95\%$ HPD interval is entirely positive). There is not sufficient evidence that there is a negative correlation for any feature pair. For the four feature pairs where the HPD interval for the correlation coefficient is entirely negative (VO-NA, VO-NNum, VS-NNum, PN-NA), the log-Bayes factors in favor of a non-zero correlation (1.60, 2.01 2.47, 3.17) are too small to merit a definite conclusion.

Conversely, for no feature pair is the Bayes factor in favor of a zero-correlation large enough to infer the absence of a correlation.\\

\section{Discussion}

\subsection{Equilibrium analysis vs.\ language sampling}

\cite{maslova2000} argues that the frequency distribution of typological feature values may be biased by accidents of history, and that the equilibrium distribution of the underlying Markov process more accurately reflects the effects of the cognitive and functional forces. Inspection of our results reveals that the difference between raw frequencies and equilibrium probabilities can be quite substantial. In Table \ref{tab:9} the relative frequencies, the equilibrium frequencies and the $95\%$ HPD intervals for the four values of the feature combination ``verb-object/adposition-noun'' are shown.
\begin{table}
 \begin{center}
  \begin{tabular}{lcccc}
   \toprule
   values  & relative frequencies & stratified relative frequencies & equilibrium (median) & HPD            \\\midrule
   OV-NAdp & 0.420                & 0.663                           & 0.614                & (0.540, 0.685) \\
   OV-AdpN & 0.011                & 0.009                           & 0.060                & (0.032, 0.096) \\
   VO-NAdp & 0.030                & 0.051                           & 0.091                & (0.054, 0.134) \\
   VO-AdpN & 0.540                & 0.278                           & 0.230                & (0.175, 0.293) \\\bottomrule
  \end{tabular}
 \end{center}
 \caption{ Relative frequencies, stratified frequencies and posterior probabilities of the four value combinations of VO-PN.}\label{tab:9}
\end{table}

We also computed the \emph{stratified frequencies}, i.e.\ the weighted means where each language is weighted by the inverse of the size of its Glottolog lineage. As a result, each lineage has the same cumulative weight.

\begin{figure}
 \begin{center}
  \includegraphics[width=8cm]{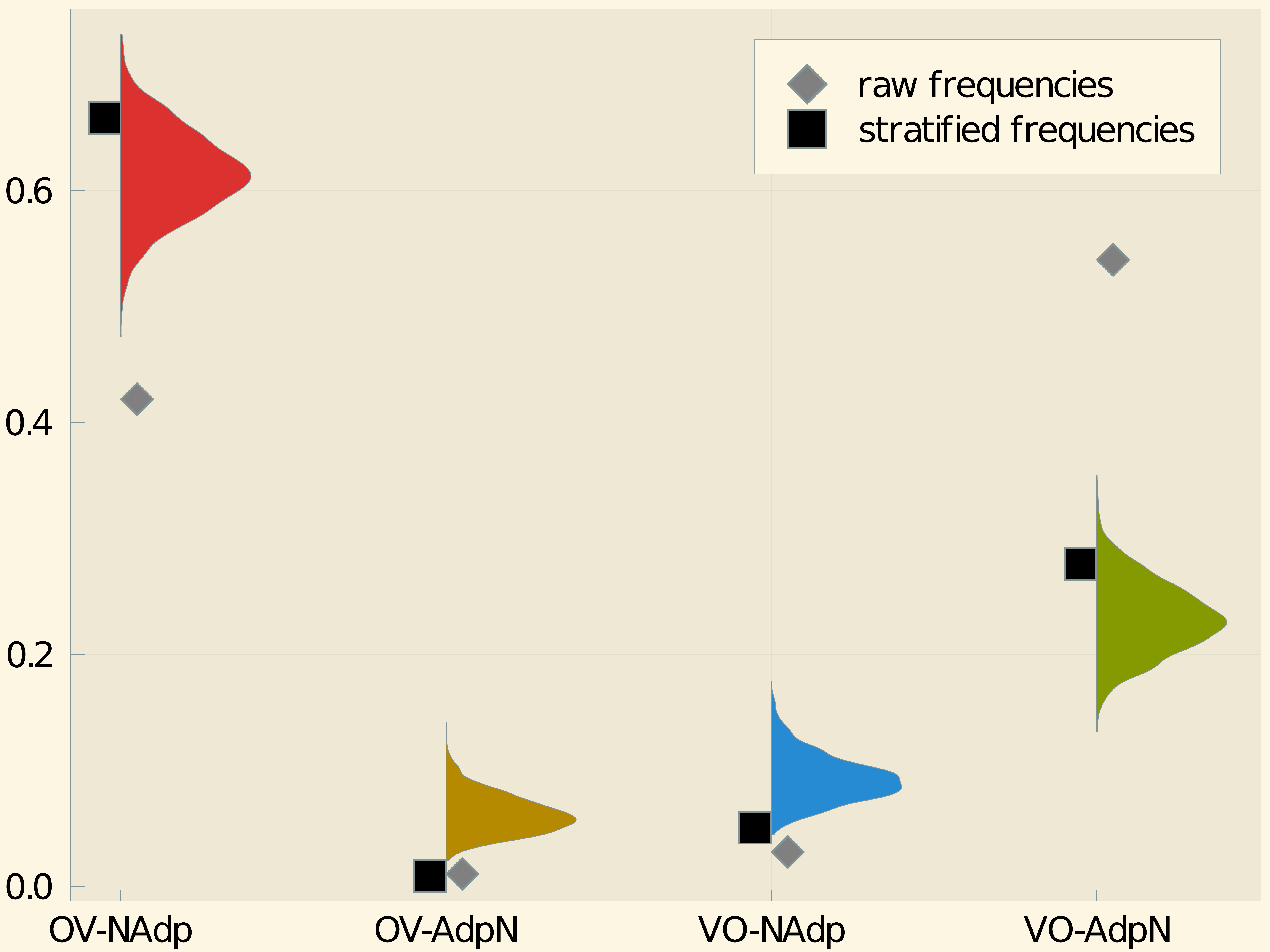}
 \end{center}
 \caption{ Relative frequencies, stratified frequencies and posterior probabilities of the four value combinations of VO-PN.}\label{fig:16}
\end{figure}

The same information is displayed in Figure \ref{fig:16}.
It can be discerned that uniformly head-initial languages (VO-AdpN) are over-represented among the languages of the world in comparison to the equilibrium distribution while uniformly head-final languages (OV-NAdp) are underrepresented. The stratified frequencies come very close to the equilibrium distribution though. This discrepancies are arguably due to the fact that head-initial languages are predominant in several large families while head-final languages are quite frequent among small families and isolates.

This example suggests that our approach effectively achieves something similar than stratified sampling, namely discounting the impact of large families and give more weight to small families and isolates. A more detailed study of the relationship between stratified sampling and equilibrium analysis is a topic for future research.\\

\subsection{Universal vs.\ lineage-specific models}

The findings from Subsection \ref{subsec:2_9} clearly demonstrate that the universal model provides a better fit of the data than the lineage-specific model. This raises the question why \cite{dunnetal11} came to the opposite conclusion. There are several relevant considerations. First, these authors did not directly test a universal model. Rather, they fitted two lineage-specific models for each feature pair --- one where the features evolve independently and one where the mutation rates of one feature may depend on the state of the other feature. They then compute the Bayes factor between these models for each family separately and conclude that the patterns of Bayes factors vary wildly between families. So essentially it is tested whether the pattern of feature correlations is identical across families.

In this paper, we explored slightly different hypotheses. We tested whether the data support a model where all lineages following the same dynamics with the same parameters (where a correlation between features is possible), or whether they support different parameters (each admitting a correlation between features). Having the same model across lineages implies an identical correlation structure, but it also implies many other things, such as identical equilibrium distributions, identical rate of change etc.

To pick an example, \cite{dunnetal11} found evidence for a correlation between NA and NRc for Austronesian and Indo-European but not for Bantu and Uto-Aztecan. This seems to speak against a universal model. However, inspection of our data reveals that the feature value ``relative clause precedes noun'' only occurs in $1.8\%$ of all Austronesian and $13.8\%$ of all Indo-European languages, and it does not occur at all in Bantu or Uto-Aztecan. The universal model correctly predicts that the observed frequency distributions will be similar across lineages (as demonstrated by the low cross-family variance in the prior predictive simulations discussed in Subsection \ref{subsec:2_6}). The lineage-specific model cannot account for this kind of cross-family similarities. More generally, our approach to test the relative merits of a universal versus lineage-specific dynamics regarding word-order features takes more sources of information into account than just correlation patterns. This more inclusive view clearly supports the universal model.\\

\subsection{Word-order correlations}

\begin{figure}
 \begin{center}
  \includegraphics[width=\linewidth]{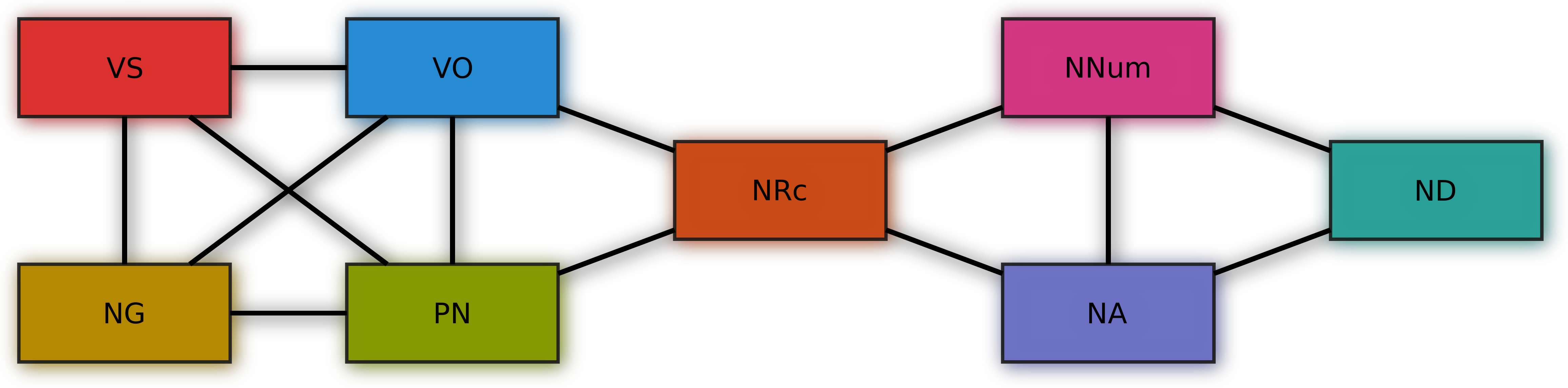}
 \end{center}
 \caption{ Feature-pairs with credible evidence for a correlation.}\label{fig:17}
\end{figure}

The thirteen feature pairs identified in Subsection \ref{subsec:2_10} for which there is credible evidence for a correlation are shown in Figure \ref{fig:17}, where connecting lines indicate credible evidence for a correlation.

The four features correlated with VO are exactly those among the features considered here that were identified by \cite{dryer92} as ``verb patterners'', i.e.\ for which he found evidence for a correlation with verb-object order. These are verb-subject, noun-genitive, adposition-noun and noun-relative clause. It is perhaps noteworthy that like Dryer, we did not find credible evidence for a correlation between verb-object order and noun-adjective order, even though such a connection has repeatedly been hypothesized, e.g.\ by \cite{lehmann73, vennemann74}, and, more recently, by \cite{ferrerLiu2014}.

Besides Dryer's verb patterns, we found a group of three mutually correlated features, noun-numeral, noun-adjective and noun-demonstrative. Two of them, noun-numeral and noun-adjective, are also correlated with noun-relative clause. These correlations have received less attention in the typological literature. The findings are not very surprising though, given that all these features pertain to the ordering of noun-phrase material relative to the head noun.\\

\section{Conclusion}

In this article we argued that the modeling of typological feature distributions in terms of phylogenetic continuous-time Markov chains --- inspired Maslova's (\citeyear{maslova2000}) theoretical work as well as by research within the framework of the biological comparative method such as \cite{pagelMeade06} and \cite{dunnetal11} --- has several advantages for typology. It allows to use all data, from families large and small as well as from isolate languages. The method controls for non-independence due to common descent. Couched in a Bayesian framework, it affords standard techniques for model checking and model comparison as well as  quantification of the uncertainty in inference. We do see it as essential though that this kind of study uses data from a variety of lineages since individual families generally do not display evidence for all the possible diachronic transitions required to estimated transition rates reliably. Working with forests rather than single trees, i.e.\ with trees or tree distributions for several families and also including isolates as elementary trees is a suitable way to achieve this goal.\footnote{\cite{verkerketal2021} use a similar approach but utilize a universal tree encompassing all lineages.}

There is a variety of open issues for future research. \cite{maslova2000} also discusses the possibility that the current distribution of feature value represents traces of proto-world or some later bottleneck language, which would bias the estimation of the equilibrium distribution. In the present paper this option was disregarded. It is possible to address this question using Bayesian model comparison.

By design, phylogenetic models only capture vertical transmission. The effects of language contact and areal tendencies are systematically ignored. In future work, this could be remedied by including areal and spatial random effects into the model.

Statistical research in other disciplines involving stratified data suggest that the binary alternative between a lineage-specific and a universal model might be ill-posed. Both approaches can be integrated within \emph{hierarchical models} (see, e.g., \citealt{bda3,mcelreath16}) where between-group variance is as small as possible but as large as the data require. Due to the high number of parameters involved, fitting such models, however, poses a considerable computational challenge.

\section*{Funding}
This research was supported by the DFG Centre for Advanced Studies in the Humanities
``Words, Bones, Genes, Tools'' (DFG-KFG 2237) and by the European Research Council (ERC)
under the European Union’s Horizon 2020 research and innovation programme (Grant agreement
834050).

\section*{Acknowledgments}

We thank Michael Franke for helpful input regarding Bayesian model comparison.

\section*{Data Availability Statement}

The code used for this study can be found at \url{https://github.com/gerhardJaeger/phylogeneticTypology}.

\setlength{\bibsep}{0pt}

\newpage
\section{Supplementary Material}

\subsection{Posterior predictive sampling}

This section contains the tables and figures reporting the results of the posterior predictive sampling (Subsection 2.8 of the main text).

\begin{figure}[htb]
 \begin{center}
  \includegraphics[width=\linewidth]{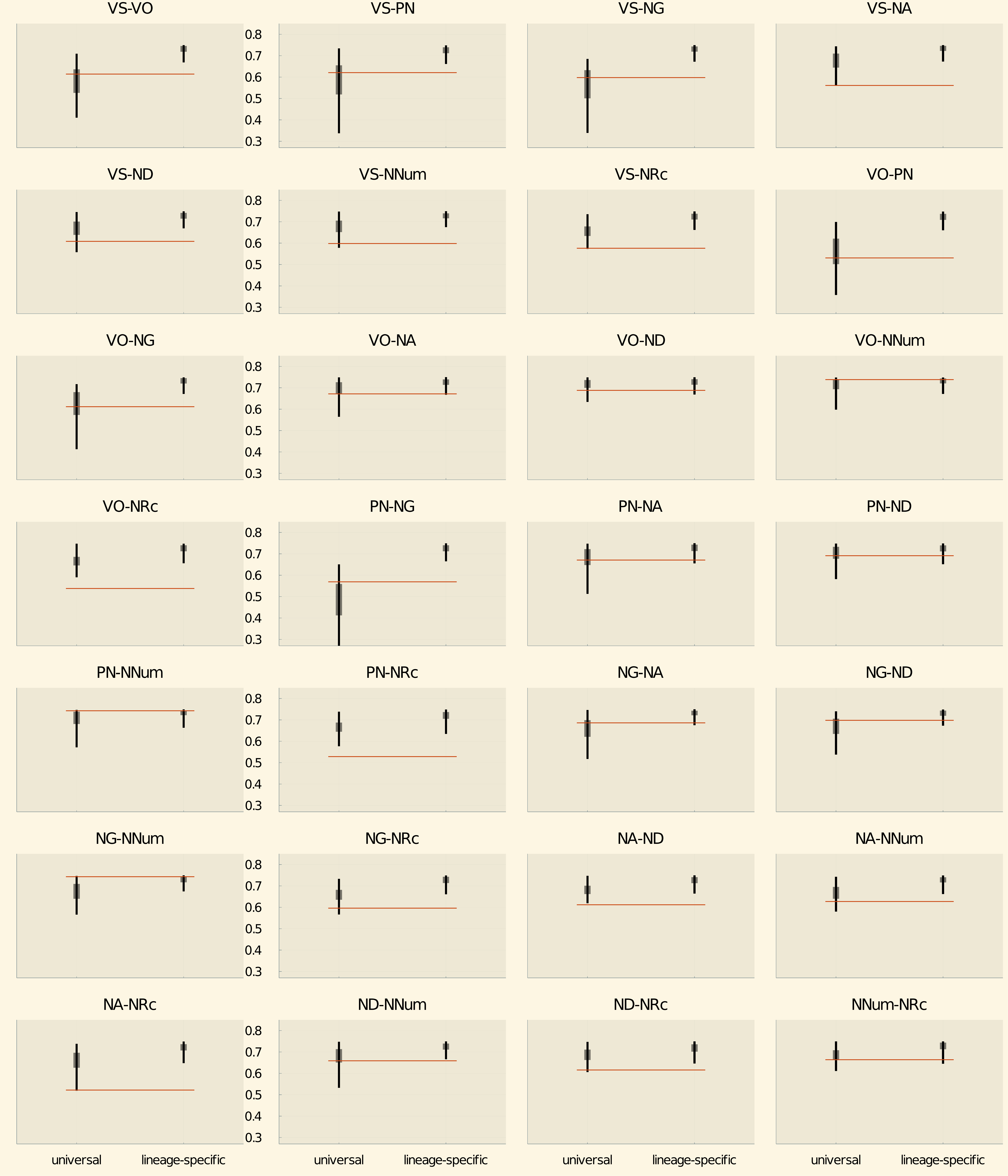}
 \end{center}
 \caption{ Posterior predictive simulations: total variance. Horizontal lines indicate the empirical value. The thick vertical lines show the $50\%$ highest-density intervals and the thin lines the $95\%$ highest-density intervals of the posterior predictive distributions.}\label{fig:8}
\end{figure}

\begin{figure}[htb]
 \begin{center}
  \includegraphics[width=\linewidth]{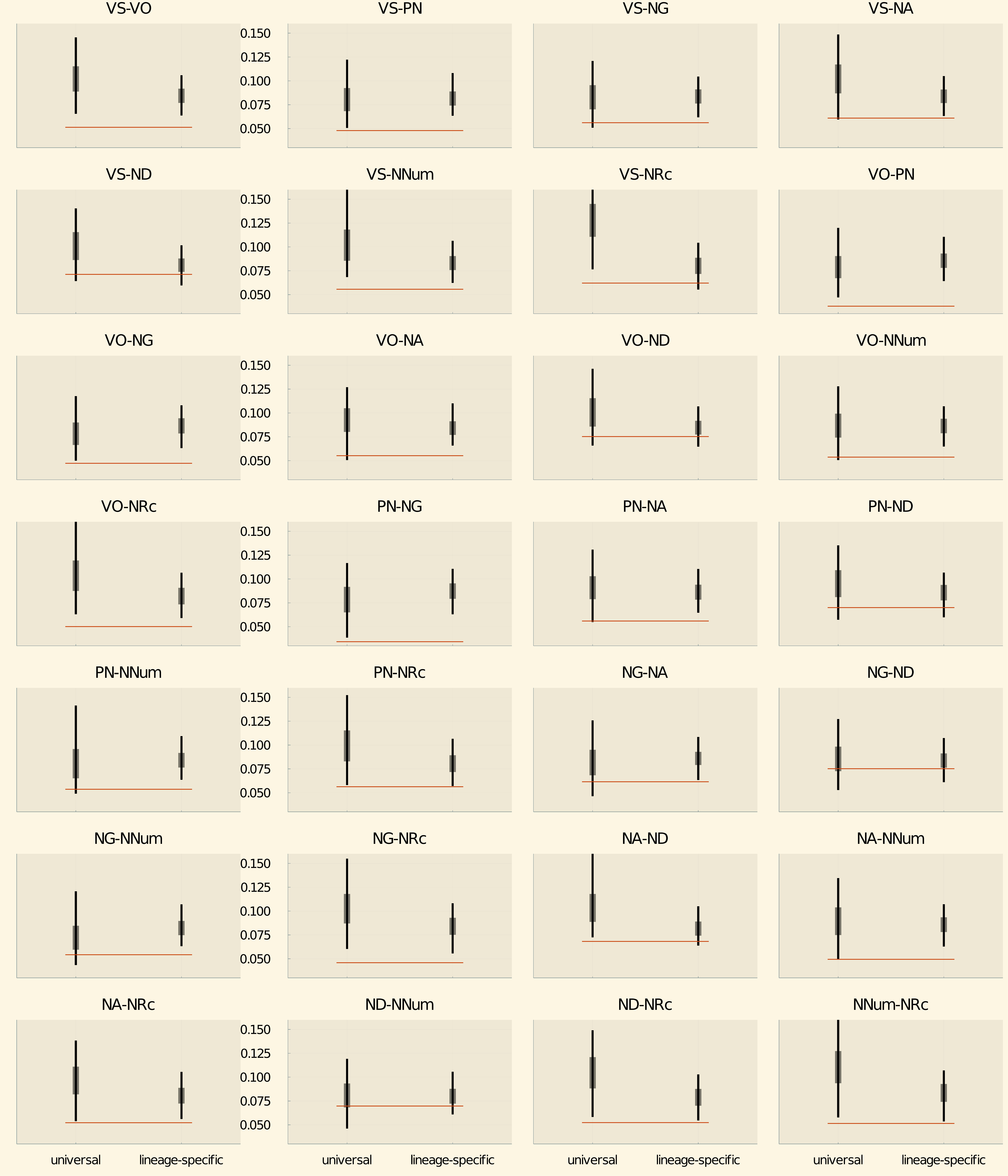}
 \end{center}
 \caption{ Posterior predictive simulations: lineage-wise variance. Horizontal lines indicate the empirical value. The thick vertical lines show the $50\%$ highest-density intervals and the thin lines the $95\%$ highest-density intervals of the posterior predictive distributions.}\label{fig:9}
\end{figure}

\begin{figure}[htp]
 \begin{center}
  \includegraphics[width=\linewidth]{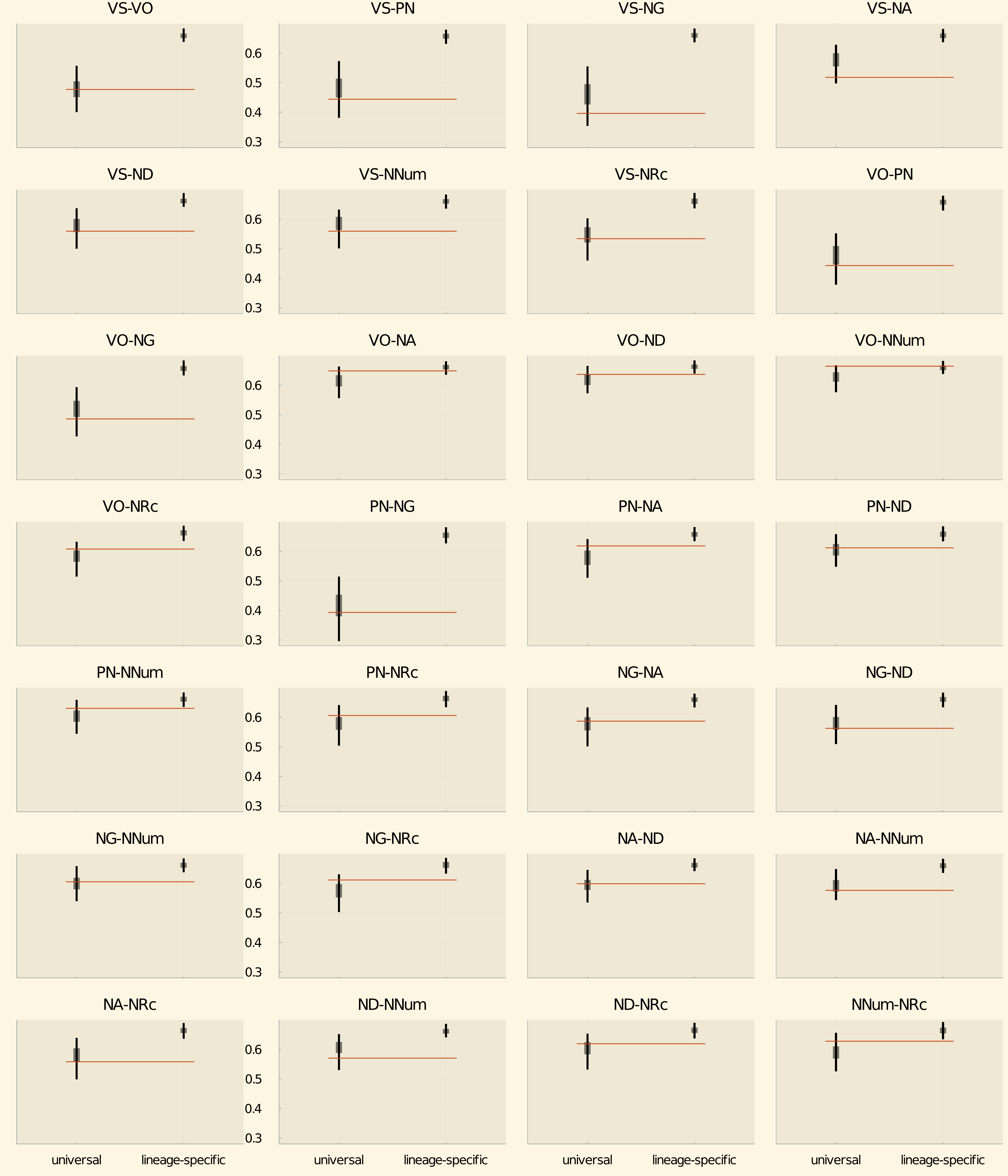}
 \end{center}
 \caption{ Posterior predictive simulations: cross-lineage variance. Horizontal lines indicate the empirical value. The thick vertical lines show the $50\%$ highest-density intervals and the thin lines the $95\%$ highest-density intervals of the posterior predictive distributions.}\label{fig:10}
\end{figure}

\begin{table}[htb]
 \begin{center}
  \begin{tabular}{lccc}
   \toprule
   feature pair & empirical & universal model (HPD)   & lineage-specific model (HPD) \\\midrule
   VS-VO        & 0.614     & \textbf{(0.410, 0.710)} & {(0.669, 0.749)}             \\
   VS-PN        & 0.621     & \textbf{(0.337, 0.734)} & {(0.661, 0.748)}             \\
   VS-NG        & 0.598     & \textbf{(0.339, 0.685)} & {(0.672, 0.749)}             \\
   VS-NA        & 0.561     & \textbf{(0.560, 0.744)} & {(0.672, 0.750)}             \\
   VS-ND        & 0.609     & \textbf{(0.558, 0.745)} & {(0.669, 0.749)}             \\
   VS-NNum      & 0.598     & \textbf{(0.578, 0.748)} & {(0.675, 0.749)}             \\
   VS-NRc       & 0.576     & \textbf{(0.573, 0.735)} & {(0.662, 0.748)}             \\
   VO-PN        & 0.531     & \textbf{(0.358, 0.699)} & {(0.660, 0.748)}             \\
   VO-NG        & 0.611     & \textbf{(0.413, 0.717)} & {(0.671, 0.748)}             \\
   VO-NA        & 0.672     & \textbf{(0.564, 0.749)} & \textbf{(0.667, 0.750)}      \\
   VO-ND        & 0.688     & \textbf{(0.634, 0.749)} & \textbf{(0.669, 0.750)}      \\
   VO-NNum      & 0.738     & \textbf{(0.598, 0.748)} & \textbf{(0.671, 0.748)}      \\
   VO-NRc       & 0.538     & {(0.590, 0.747)}        & {(0.656, 0.748)}             \\
   PN-NG        & 0.569     & \textbf{(0.270, 0.651)} & {(0.665, 0.749)}             \\
   PN-NA        & 0.671     & \textbf{(0.513, 0.747)} & \textbf{(0.656, 0.750)}      \\
   PN-ND        & 0.691     & \textbf{(0.582, 0.748)} & \textbf{(0.651, 0.749)}      \\
   PN-NNum      & 0.742     & \textbf{(0.572, 0.747)} & \textbf{(0.663, 0.749)}      \\
   PN-NRc       & 0.528     & {(0.577, 0.739)}        & {(0.634, 0.748)}             \\
   NG-NA        & 0.686     & \textbf{(0.518, 0.746)} & \textbf{(0.675, 0.749)}      \\
   NG-ND        & 0.698     & \textbf{(0.537, 0.740)} & \textbf{(0.673, 0.748)}      \\
   NG-NNum      & 0.743     & \textbf{(0.566, 0.747)} & \textbf{(0.674, 0.750)}      \\
   NG-NRc       & 0.596     & \textbf{(0.566, 0.733)} & {(0.661, 0.749)}             \\
   NA-ND        & 0.612     & {(0.619, 0.747)}        & {(0.664, 0.749)}             \\
   NA-NNum      & 0.627     & \textbf{(0.580, 0.743)} & {(0.662, 0.749)}             \\
   NA-NRc       & 0.522     & \textbf{(0.518, 0.738)} & {(0.647, 0.749)}             \\
   ND-NNum      & 0.659     & \textbf{(0.532, 0.748)} & {(0.666, 0.750)}             \\
   ND-NRc       & 0.616     & \textbf{(0.606, 0.747)} & {(0.646, 0.749)}             \\
   NNum-NRc     & 0.664     & \textbf{(0.611, 0.750)} & \textbf{(0.645, 0.749)}      \\\bottomrule
  \end{tabular}
 \end{center}
 \caption{Posterior predictive simulation: Total variance. Intervals are shown in bold if
  they include the empirical value.}\label{tab:2}
\end{table}

\begin{table}[htb]
 \begin{center}
  \begin{tabular}{lccc}
   \toprule
   feature pair & empirical & universal model (HPD)   & lineage-specific model (HPD) \\\midrule
   VS-VO        & 0.051     & {(0.066, 0.146)}        & {(0.064, 0.106)}             \\
   VS-PN        & 0.048     & {(0.051, 0.122)}        & {(0.063, 0.108)}             \\
   VS-NG        & 0.056     & \textbf{(0.051, 0.121)} & {(0.062, 0.104)}             \\
   VS-NA        & 0.061     & \textbf{(0.060, 0.149)} & {(0.063, 0.105)}             \\
   VS-ND        & 0.071     & \textbf{(0.064, 0.140)} & \textbf{(0.060, 0.102)}      \\
   VS-NNum      & 0.056     & {(0.068, 0.163)}        & {(0.062, 0.106)}             \\
   VS-NRc       & 0.062     & {(0.077, 0.183)}        & \textbf{(0.055, 0.104)}      \\
   VO-PN        & 0.038     & {(0.047, 0.120)}        & {(0.064, 0.111)}             \\
   VO-NG        & 0.047     & {(0.050, 0.118)}        & {(0.063, 0.108)}             \\
   VO-NA        & 0.055     & \textbf{(0.051, 0.127)} & {(0.066, 0.110)}             \\
   VO-ND        & 0.075     & \textbf{(0.066, 0.146)} & \textbf{(0.065, 0.107)}      \\
   VO-NNum      & 0.054     & \textbf{(0.051, 0.128)} & {(0.065, 0.107)}             \\
   VO-NRc       & 0.050     & {(0.063, 0.163)}        & {(0.059, 0.107)}             \\
   PN-NG        & 0.034     & {(0.038, 0.117)}        & {(0.063, 0.111)}             \\
   PN-NA        & 0.056     & \textbf{(0.055, 0.131)} & {(0.065, 0.110)}             \\
   PN-ND        & 0.070     & \textbf{(0.057, 0.135)} & {(0.060, 0.107)}             \\
   PN-NNum      & 0.054     & \textbf{(0.049, 0.141)} & {(0.064, 0.109)}             \\
   PN-NRc       & 0.056     & {(0.058, 0.152)}        & {(0.057, 0.107)}             \\
   NG-NA        & 0.062     & \textbf{(0.046, 0.126)} & {(0.063, 0.108)}             \\
   NG-ND        & 0.075     & \textbf{(0.053, 0.127)} & \textbf{(0.061, 0.107)}      \\
   NG-NNum      & 0.054     & \textbf{(0.043, 0.121)} & {(0.063, 0.107)}             \\
   NG-NRc       & 0.046     & {(0.060, 0.155)}        & {(0.056, 0.108)}             \\
   NA-ND        & 0.068     & {(0.072, 0.164)}        & \textbf{(0.064, 0.105)}      \\
   NA-NNum      & 0.049     & \textbf{(0.049, 0.135)} & {(0.063, 0.107)}             \\
   NA-NRc       & 0.052     & {(0.054, 0.138)}        & {(0.056, 0.105)}             \\
   ND-NNum      & 0.070     & \textbf{(0.046, 0.119)} & \textbf{(0.061, 0.106)}      \\
   ND-NRc       & 0.052     & {(0.058, 0.149)}        & {(0.055, 0.103)}             \\
   NNum-NRc     & 0.051     & {(0.058, 0.162)}        & {(0.054, 0.107)}             \\\bottomrule
  \end{tabular}
 \end{center}
 \caption{Posterior predictive simulation: Lineage-wise variance. Intervals are shown in
  bold if they include the empirical value.}\label{tab:3}
\end{table}

\begin{table}[htb]
 \begin{center}
  \begin{tabular}{lccc}
   \toprule
   feature pair & empirical & universal model (HPD)   & lineage-specific model (HPD) \\\midrule
   VS-VO        & 0.477     & \textbf{(0.400, 0.557)} & {(0.638, 0.684)}             \\
   VS-PN        & 0.444     & \textbf{(0.381, 0.573)} & {(0.631, 0.680)}             \\
   VS-NG        & 0.396     & \textbf{(0.354, 0.555)} & {(0.637, 0.683)}             \\
   VS-NA        & 0.518     & \textbf{(0.497, 0.629)} & {(0.637, 0.682)}             \\
   VS-ND        & 0.559     & \textbf{(0.500, 0.637)} & {(0.642, 0.688)}             \\
   VS-NNum      & 0.559     & \textbf{(0.501, 0.632)} & {(0.636, 0.684)}             \\
   VS-NRc       & 0.534     & \textbf{(0.460, 0.603)} & {(0.636, 0.689)}             \\
   VO-PN        & 0.443     & \textbf{(0.378, 0.552)} & {(0.629, 0.679)}             \\
   VO-NG        & 0.486     & \textbf{(0.427, 0.593)} & {(0.633, 0.684)}             \\
   VO-NA        & 0.649     & \textbf{(0.556, 0.663)} & \textbf{(0.636, 0.681)}      \\
   VO-ND        & 0.637     & \textbf{(0.572, 0.666)} & {(0.640, 0.684)}             \\
   VO-NNum      & 0.664     & \textbf{(0.576, 0.668)} & \textbf{(0.638, 0.683)}      \\
   VO-NRc       & 0.607     & \textbf{(0.514, 0.632)} & {(0.635, 0.687)}             \\
   PN-NG        & 0.393     & \textbf{(0.296, 0.515)} & {(0.627, 0.682)}             \\
   PN-NA        & 0.618     & \textbf{(0.510, 0.641)} & {(0.634, 0.682)}             \\
   PN-ND        & 0.612     & \textbf{(0.548, 0.658)} & {(0.633, 0.685)}             \\
   PN-NNum      & 0.630     & \textbf{(0.544, 0.659)} & {(0.635, 0.684)}             \\
   PN-NRc       & 0.606     & \textbf{(0.504, 0.642)} & {(0.634, 0.689)}             \\
   NG-NA        & 0.587     & \textbf{(0.502, 0.634)} & {(0.633, 0.680)}             \\
   NG-ND        & 0.563     & \textbf{(0.509, 0.642)} & {(0.634, 0.684)}             \\
   NG-NNum      & 0.605     & \textbf{(0.540, 0.659)} & {(0.637, 0.685)}             \\
   NG-NRc       & 0.612     & \textbf{(0.503, 0.631)} & {(0.633, 0.686)}             \\
   NA-ND        & 0.599     & \textbf{(0.535, 0.646)} & {(0.641, 0.685)}             \\
   NA-NNum      & 0.576     & \textbf{(0.543, 0.648)} & {(0.636, 0.683)}             \\
   NA-NRc       & 0.558     & \textbf{(0.499, 0.639)} & {(0.636, 0.689)}             \\
   ND-NNum      & 0.571     & \textbf{(0.530, 0.652)} & {(0.641, 0.687)}             \\
   ND-NRc       & 0.619     & \textbf{(0.532, 0.653)} & {(0.637, 0.690)}             \\
   NNum-NRc     & 0.628     & \textbf{(0.526, 0.656)} & {(0.634, 0.693)}             \\\bottomrule
  \end{tabular}
 \end{center}
 \caption{Posterior predictive simulation: Cross-lineage variance. Intervals are shown in
  bold if they include the empirical value.}\label{tab:4}
\end{table}

\subsection{Transition rates: posterior distributions}

\begin{figure}
  \begin{center}
    \includegraphics[width=.7\linewidth]{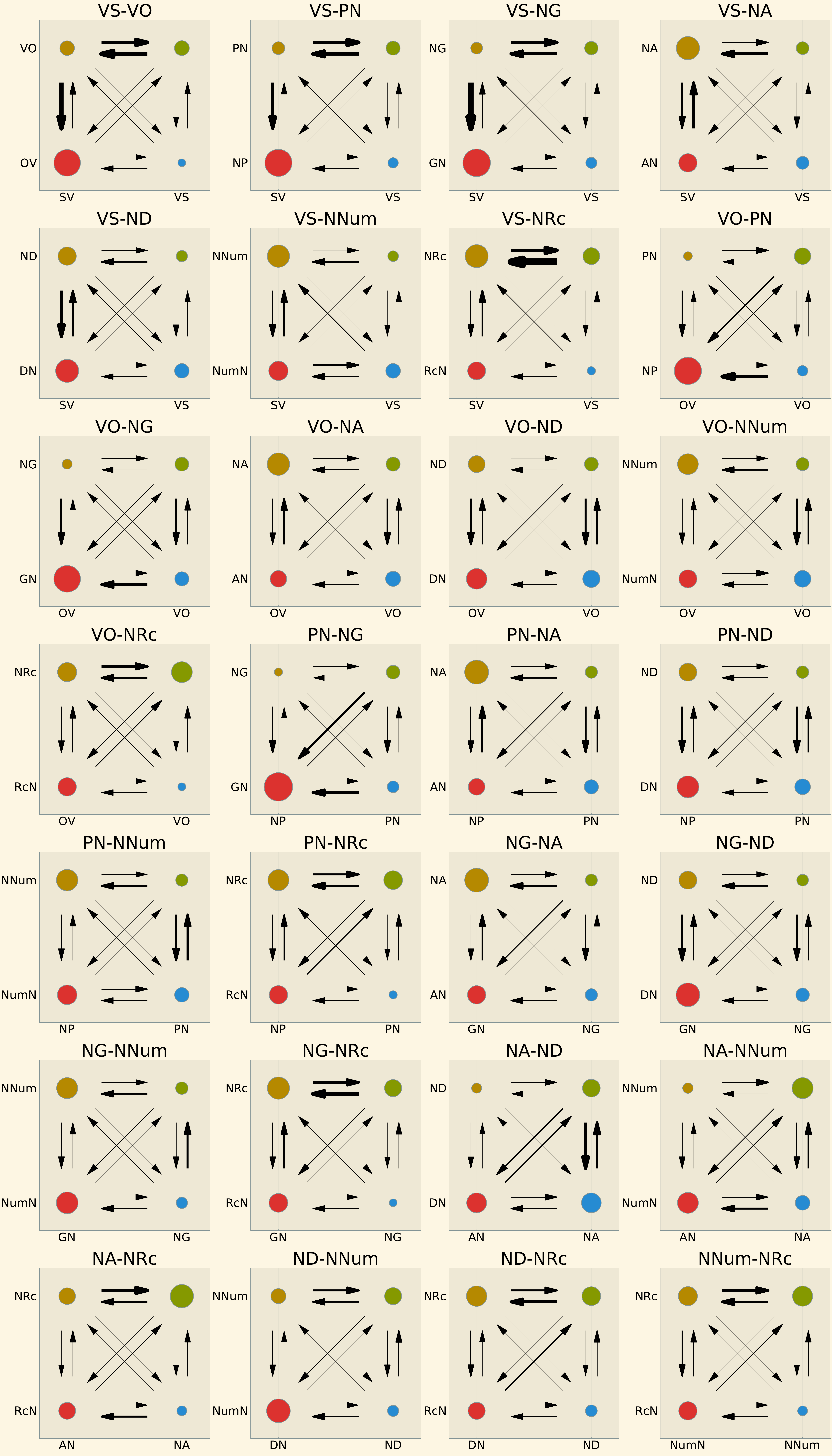}
  \end{center}
  \caption{Fitted CTMC parameters. Thickness of the arrows represents the median posterior values of the corresponding rates, and area of the circles is proportional to the median equilibrium distributions of the corresponding states.}\label{fig:18}
\end{figure}

A visualization of the magnitude of the transition rates for all feature pairs is shown in Figure \ref{fig:18}.

The medians and HPD intervals can be seen in \url{https://github.com/gerhardJaeger/phylogeneticTypology/blob/main/data/posteriorStatistics.csv}.

\end{document}